\documentclass[pdflatex,sn-mathphys-num]{sn-jnl}

\usepackage{graphicx}%
\usepackage{multirow}%
\usepackage{rotating}%
\usepackage{amsmath,amssymb,amsfonts}%
\usepackage{amsthm}%
\usepackage{mathrsfs}%
\usepackage[title]{appendix}%
\usepackage{xcolor}%
\usepackage{textcomp}%
\usepackage{manyfoot}%
\usepackage{booktabs}%
\usepackage{algorithm}%
\usepackage{algorithmicx}%
\usepackage{algpseudocode}%
\usepackage{listings}%

\theoremstyle{thmstyleone}%
\theoremstyle{thmstyletwo}%

\theoremstyle{thmstylethree}%

\begin{document}

\title[Article Title]{The Cost of Consensus: Malignant Epistemic Herding and Adaptive Gating in Distributed Multi-Agent Search}


\author*[1]{\fnm{David} \sur{Farr}}\email{dtfarr@uw.edu}
\author[2]{\fnm{Iain} \sur{Cruickshank}}\email{icruicks@andrew.cmu.edu}
\author[3]{\fnm{Kate Starbird} \sur{}}\email{kstarbi@uw.edu}
\author[1]{\fnm{Jevin} \sur{West}}\email{jevinw@uw.edu}

\affil*[1]{\orgdiv{Information School}, \orgname{University of Washington}, \orgaddress{\street{1410 NE Campus Parkway}, \city{Seattle}, \postcode{98195}, \state{WA}, \country{USA}}}

\affil[2]{\orgdiv{Computer Science}, \orgname{Carnegie Mellon University}, \orgaddress{\street{5000 Forbes Avenue}, \city{Pittsburgh}, \postcode{15213}, \state{PA}, \country{USA}}}

\affil[3]{\orgdiv{Human Centered Design Engineering}, \orgname{University of Washington}, \orgaddress{\street{1410 NE Campus Parkway}, \city{Seattle}, \postcode{98195}, \state{WA}, \country{USA}}}


\abstract{Distributed multi-agent systems are increasingly leveraged in conditions of partial observability and high uncertainty where communication and coordination is critical to task success, yet the relationship between communication design and collective belief quality remains poorly understood. We investigate how message content and frequency jointly shape both task performance and epistemic alignment, the degree to which agents share consistent probabilistic beliefs about the environment. We operationalize this through a decentralized multi-agent search task with a full factorial simulation study (108 conditions, 1,000 episodes per condition, 
T=200 timesteps per episode). Epistemic messages that transmit compressed belief distributions substantially improve task success over point-estimate alternatives, yet high-frequency epistemic communication paradoxically degrades collective belief quality by driving belief convergence through repeated fusion, a failure mode we formalize as malignant epistemic herding (MEH), a form of coordinated belief convergence analogous to herding behavior but arising through repeated belief fusion rather than social action imitation, and distinguished by its undetectability in standard coordination metrics. To resolve this tension, we propose entropy-delta gating, an adaptive mechanism that conditions transmission on information novelty using 
Shannon entropy as both a transmission gate and a fusion weight. 
Entropy-delta gating reduces message volume by over 98\% relative to 
ungated epistemic communication while achieving the highest alignment 
to truth and lowest MEH rate across all coordination 
thresholds. Our results establish MEH as a team-level 
failure mode distinct from individual overconfidence, and entropy-delta 
gating as a lightweight, self-regulating coordination primitive 
applicable to any decentralized system in which agents maintain 
probabilistic beliefs under uncertainty.}




\maketitle

\section{INTRODUCTION}

Distributed agents in real-world settings frequently must coordinate under uncertainty with only partial 
observations. Coordination is necessary to share beliefs to aid in task completion, but communication 
costs bandwidth, introduces latency, and if done poorly, can degrade 
collective reasoning. This tension is especially acute in bandwidth-constrained deployments such as distributed sensing networks, autonomous reconnaissance, and collaborative cyber defense, where excessive transmission carries direct operational costs. Existing work has focused on multi-agent exploration and communication strategies, but not on how communication frequency and content jointly shape the collective belief state.

Central to this challenge is the degree to which agents maintain compatible internal beliefs about the environment, a property we term \textit{epistemic alignment}. When agents share beliefs effectively, they converge on correct hypotheses; when communication is poorly designed, agents may converge confidently on wrong ones. We formalize this distinction and show it is not detectable from coordination metrics alone such as Jensen-Shannon Divergence or rate to consensus.

In this work we make the following contributions:

\begin{itemize}
    \item \textbf{Conceptual:} We formalize \textit{epistemic alignment} as a property of distributed belief systems and introduce 
    \textit{malignant epistemic herding (MEH)}, a previously underexplored failure mode in which agents converge to a shared but incorrect belief through repeated belief fusion rather than social action imitation, rendering failure undetectable in standard coordination metrics.

    \item \textbf{Methodological:} We propose entropy-delta gating, an adaptive communication mechanism that conditions transmission on information novelty, jointly regulating communication frequency and influence.

    \item \textbf{Empirical:} We evaluate four communication protocols ($C_0$--$C_3$) under a full factorial simulation design spanning communication type, packet loss, latency, and coordination requirements, demonstrating that communication frequency and message content drive distinct aspects of collective performance.

    \item \textbf{Systems Insight:} We show that high-frequency epistemic communication degrades epistemic health through self-reinforcing belief convergence, an effect mitigated by adaptive gating without sacrificing task success.
\end{itemize}

Our central finding is that message content and message frequency drive distinct aspects of collaborative agent performance. Epistemic messages (e.g., a message where an agent shares their belief distribution) improve task success rates substantially over point-estimate alternatives, yet high-frequency epistemic communication paradoxically worsens epistemic health by driving belief convergence through repeated fusion. Entropy-delta gating resolves this tension: by transmitting only when belief has shifted significantly, 
$C_3$ reduces message volume by over 98\% while achieving the highest epistemic alignment and lowest MEH rate across all coordination thresholds.

The remainder of the paper is organized as follows. Section 2 reviews related work on multi-agent communication, decentralized search, and belief-based coordination. Section 3 describes the simulation environment, agent architecture, and the four communication protocols. Section 4 defines our performance and epistemic-alignment metrics and the experimental design. Section 5 presents results for each research question. Section 6 interprets the findings and discusses limitations.

\section{Related Work}
\label{sec:related}
Our work sits at the intersection of four bodies of literature: decentralized multi-agent search and coordination, distributed Bayesian belief fusion, event-triggered and adaptive communication, 
and information-theoretic sensing. We use the term \textit{agent} throughout in place of \textit{robot}, as the target applications include software agents operating over computer networks in addition to physical robotic systems.

\subsection{Decentralized multi-agent search and coordination}

Decentralized coordination in multi-agent systems has been studied extensively under the framework of cooperative task allocation and fault-tolerant teaming. \cite{parker2002alliance}'s ALLIANCE architecture demonstrated that agents can coordinate through motivational behavior sets without centralized control, but does not address how shared beliefs should be encoded. \citet{stone2000multiagent} survey coordination strategies from a machine learning perspective, noting that communication design is a first-class concern in cooperative settings. Our work extends this line of inquiry by isolating the encoding of belief content as the key independent variable, rather than scheduling of communication or introducing a centralized orchestrating agent.

Information-theoretic approaches have informed both agent movement and sensing in multi-robot systems. Charrow~\cite{charrow2015information} and Frew~\cite{frew2009information} demonstrate that planners which maximize expected reduction in map entropy outperform reactive baselines in probabilistic 
search, guiding agents toward regions that most reduce collective uncertainty. Grocholsky et al.~\cite{grocholsky2006cooperative} extend this to decentralized teams, showing that information-theoretic objectives can be decomposed across agents without centralized coordination. 
Our work is complementary: rather than optimizing movement to reduce uncertainty, we ask how the communication of that uncertainty affects collective belief quality. We deliberately use a greedy movement policy to isolate communication encoding as the experimental 
variable, independent of planning quality.

\subsection{Belief Fusion and Distributed Bayesian Methods}

In decentralized multi-agent systems, each agent maintains a local 
belief and updates it using periodic messages from peers. Treating 
incoming messages as independent new evidence introduces a well-known 
failure mode called \textit{overconfidence} \cite{julier1997new}: 
because agents share information over time, fusing a peer's message 
without accounting for shared history causes evidence to be counted 
multiple times, producing beliefs that are more certain than the 
underlying observations warrant \cite{julier1997new, abu2017distributed}.

\citet{julier1997new} address this with Covariance Intersection (CI), 
which assumes incoming estimates are maximally correlated and weights 
each by its inverse uncertainty. More confident estimates exert greater 
influence on the fused result, while uncertain estimates are 
down-weighted, preventing any single agent from dominating the shared 
belief before its evidence has been corroborated. Our $C_2$ and $C_3$ 
protocols apply this same principle to discrete probability distributions over grid cells, operationalizing uncertainty through Shannon entropy 
rather than covariance matrices. Shannon entropy $H(b) = -\sum_{c \in \mathcal{S}} b(c) \log b(c)$ provides a natural measure of belief uncertainty~\cite{shannon1948mathematical}, ranging from $\log |\mathcal{S}|$ nats for a uniform distribution to zero for a deterministic belief.

Our work differs from prior distributed Bayesian fusion methods in three important ways. First, CI operates over Gaussian distributions where cross-correlation bounds can be derived analytically. Our agents maintain discrete probability distributions over grid cells, for which these analytical guarantees do not extend. Exact fusion would require full knowledge of the joint correlation structure across all agents,
unavailable in decentralized systems by definition, making our 
inverse-entropy weighting an explicitly approximate method inspired by CI but adapted to a discrete setting. Second, prior work typically assumes agents transmit full observations or likelihood functions to a central fusion mechanism; our agents transmit compressed top-$k$ summaries over channels subject to packet loss and latency. Third, while overconfidence is well characterized analytically in the CI literature, its consequences for team-level task performance have received limited empirical study. We address this gap by introducing \emph{MEH}, low inter-agent divergence combined with poor alignment to ground truth, a particularly dangerous failure mode in which agents appear coordinated while collectively converging on an incorrect solution.

\subsection{Event-Triggered and Adaptive Communication}

Event-triggered communication reduces unnecessary transmissions by 
only allowing agents to broadcast when a trigger condition is 
satisfied, and has been shown to effectively reduce message volume 
in multi-agent settings \cite{nowzari2019event}. Our $C_3$ protocol 
belongs to this family but differs in how the trigger condition is 
defined. Event-triggered messaging stems from control theory and 
traditionally fires when a state deviates from an expected norm. 
Our $C_3$ protocol instead transmits when the change in an agent's 
Shannon entropy exceeds a threshold $\theta$:
\begin{equation}
    |H(t) - H(t_{\text{last}})| \geq \theta
\end{equation}
This is interpretable as a measure of information novelty: an agent 
transmits only when its belief has shifted significantly in a given timestep, indicating 
new evidence worth sharing. Entropy also serves a second role in our 
methodology --- the absolute entropy of the sender determines how 
much influence a transmitted message carries to receiving agents 
under our inverse-entropy fusion rule. These two mechanisms 
complement each other: a discovery event causes entropy to drop 
sharply, simultaneously triggering a transmission and increasing 
the weight that transmission carries at the receiver.

The foundational result for event-triggered consensus in multi-agent 
systems is due to Dimarogonas et al.~\cite{dimarogonas2011distributed}, who 
showed that state-deviation triggers can achieve consensus with 
substantially fewer transmissions than periodic schemes while 
preserving convergence guarantees. Strom and 
Bernhardsson~\cite{astrom2002comparison} provide the theoretical basis for 
why event-triggered sampling outperforms periodic sampling in 
stochastic settings --- a result directly relevant to our setting 
where belief updates are driven by noisy observations. Our $C_3$ 
protocol differs from this family in that the trigger operates 
over a full probability distribution using Shannon entropy as a 
summary statistic, rather than over a lower-dimensional state 
vector. This distinction means $C_3$'s trigger condition captures 
the \textit{information content} of the agent's entire belief 
state in a single scalar computation, and that the same quantity 
that triggers transmission also determines message influence at 
the receiver --- a dual role that, to our knowledge, has not been 
previously characterized in the event-triggered literature.

\subsection{Learned Communication Protocols}

A similar line of research has focused on end-to-end learned communications protocols in which agents jointly optimize what to transmit and when through multi-agent reinforcement learning. Sukhbaatar et al.~\cite{sukhbaatar2016learning} introduced CommNet, demonstrating that continuous communication vectors learned via backpropagation enable effective coordination in cooperative tasks. Using reinforcement learning methods, Foerster et al.~\cite{foerster2016learning} showed that agents can learn \textit{when} to communicate as well as \textit{what} to transmit, producing sparse communication schedules that emerge from training rather than hand-designed rules. Lowe et al ~\cite{lowe2017multi} extend this to mixed cooperative-competitive settings, demonstrating that learned protocols can adapt to adversarial pressure that hand designed protocols cannot anticipate.

Our work takes a complementary approach. Hand-designed protocols offer interpretability and reproducibility that learned protocols do not, properties which are critical in high-stakes deployments where communication failures must be diagnosable. Hand-designed protocols also allow us to isolate communication properties such as content richness, transmission frequency, and information novelty as independent experimental variables, which is essential for understanding the mechanisms that drive epistemic alignment and MEH. Learned protocols trained to maximize task success would likely suppress epistemic herding implicitly without revealing why, our approach makes the mechanism visible. We view learned protocols as a natural extension and suggest in Section~\ref{future} that they could serve as an upper bound on achievable coordination efficiency against which hand-designed protocols can be benchmarked.

\section{System Model}
\label{sec:model}

\begin{figure}[t]
\centering
\fbox{\includegraphics[width=\columnwidth]{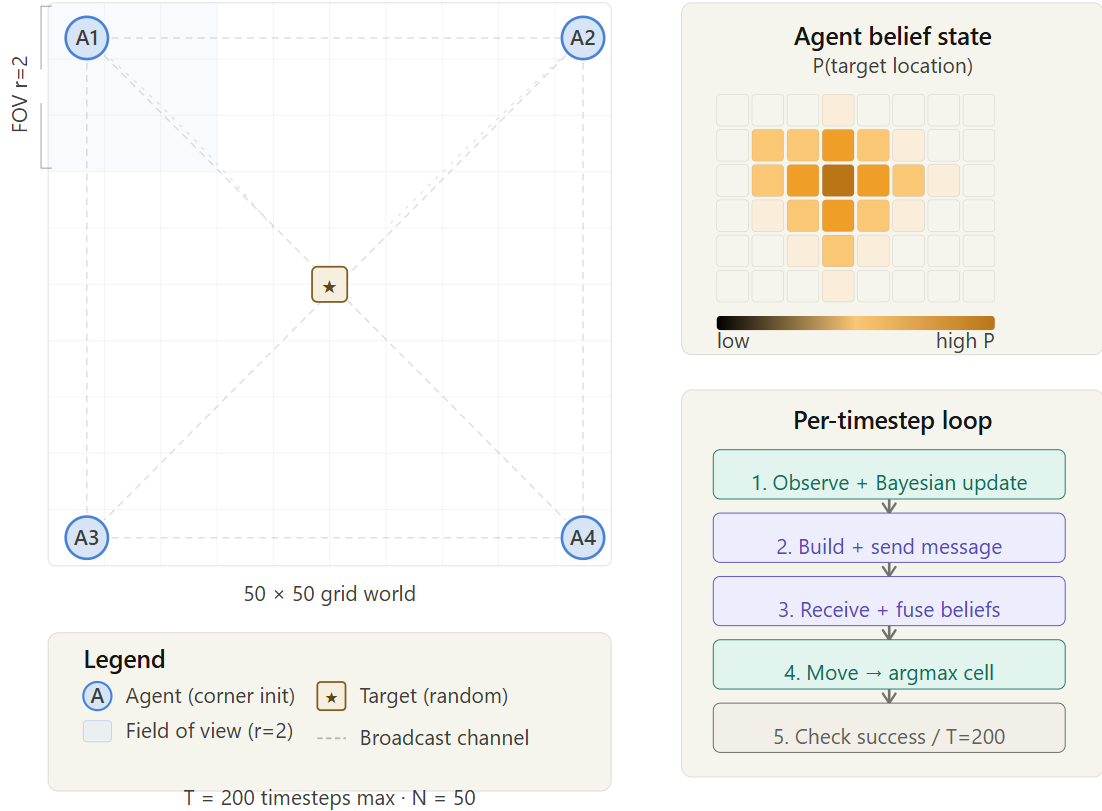}}
\caption{Simulation environment overview. \textit{Left}: the $50 \times 50$
grid world with four agents initialized at corners (A1--A4), a randomly placed target, and fully connected broadcast communication
links (dashed). The shaded region illustrates one agent's field of view ($r = 2$, Chebyshev distance). \textit{Top right}: a representative agent belief state $b_i \in \Delta(\mathcal{S})$ over grid cells; warm colors indicate higher probability mass. \textit{Bottom right}: the synchronous per-timestep execution loop applied to all agents before any fusions are resolved.}
\label{fig:system_overview}
\end{figure}

We model the search task as a Dec-POMDP \cite{oliehook}, in which $n = 4$ agents operate over a shared discrete state space $\mathcal{S}$ of $N^2$ grid cells with $N = 50$, each receiving local observations of cells within a Chebyshev-distance radius $r = 2$. Agents share a cooperative objective: at least $k \in \{2, 3, 4\}$ agents must reach the true target cell $c^* \in \mathcal{S}$, placed uniformly at random at episode start, within $T = 200$ timesteps. No agent has prior knowledge of $c^*$; the only path to coordination is through the communication protocol, which we treat as the primary experimental variable across four designs ($C_0$ -- $C_3$). Figure \ref{fig:system_overview} shows a system diagram of the multi-agent simulation and Table \ref{tab:params} shows the model parameters tested.

\begin{table}[t]
\centering
\caption{Simulation parameters used in all reported experiments unless
         otherwise noted.}
\label{tab:params}
\begin{tabular}{llr}
\toprule
Parameter & Description & Value \\
\midrule
\multicolumn{3}{l}{\textit{Environment}} \\
$N$               & Grid side length                    & 50           \\
$|\mathcal{S}|$   & State space size                    & 2{,}500 cells (50 x 50 grid) \\
$n$               & Number of agents                    & 4            \\
$r$               & Field-of-view radius (Chebyshev)    & 2 cells      \\
$T$               & Maximum timesteps per episode       & 200          \\
$k$               & Min.\ agents required for success   & $\{2, 3, 4\}$ \\
\midrule
\multicolumn{3}{l}{\textit{Communication ($C_2$ and $C_3$)}} \\
$k_{\text{msg}}$ & Top-$k$ cells transmitted per message & 5 \\
$w_{\max}$ & Maximum fusion weight per sender & 0.8 \\
$\theta$ & Entropy-delta gate threshold ($C_3$ only) & 0.20 nats \\
\midrule
\multicolumn{3}{l}{\textit{Network}} \\
$p_{\text{base}}$ & Base packet loss rate & $\{0.0, 0.1, 0.3\}$ \\
$\ell$ & Message latency (timesteps) & $\{0, 1, 3\}$ \\
\midrule
\multicolumn{3}{l}{\textit{Experiment}} \\
& Episodes per condition & 1,000 \\
& Total conditions & 108 \\
\bottomrule
\end{tabular}
\end{table}

\subsection{Simulation Environment}

The grid world is a discrete $N \times N$ bounded space in which agents navigate by moving one cell per timestep in any of the four cardinal directions. The single target $c^*$ is stationary throughout each episode and unknown to all agents at initialization. Agents maintain no shared map and carry no memory beyond their current belief distribution $b_i$.

Each agent observes a $(2r+1)^2 = 25$-cell patch per timestep, but moving one cell per step yields approximately 5 novel cells per step during linear traversal, meaning a single agent can cover at most $5 \times T = 1{,}000$ of the $2{,}500$ cells within the episode budget, roughly 40\% of the state space. This structural coverage deficit makes inter-agent coordination consequential: protocols that cause agents to redundantly search the same region directly reduce the team's effective coverage and increase the probability of exhausting the $T = 200$ step budget before locating $c^*$.

\subsection{Agent Architecture}

Each agent maintains a probabilistic belief over the state space 
$\mathcal{S}$, updates that belief from local observations and peer 
messages, and moves greedily toward its current maximum-probability cell. Agents are homogeneous, identical in sensor capability, belief representation, and movement policy, differing only in position and the communication protocol under which they operate.

\subsubsection{Belief Update}

Each agent maintains a log-probability vector over all $N^2$ cells, 
initialized to a uniform prior with small random perturbations to break argmax ties during early exploration. We denote this belief state $b_i$.

At each timestep the belief is updated in two stages. First, the agent incorporates its own sensor observations: cells observed without a detection have their log-probability decreased; the target cell, once within FOV, receives a large positive update. Second, if the agent receives peer messages, it fuses them using the inverse-entropy weighting rule described in  Section~\ref{sec:protocols}, more confident senders contribute more to the update than uncertain ones.

All arithmetic is performed in log space to prevent numerical underflow over the course of a $T = 200$ step episode \cite{thrun2002probabilistic}. Following each update, 
log-beliefs are exponentiated and renormalized so that $b_i$ remains a valid probability distribution over $\mathcal{S}$.

\subsubsection{Movement Policy}

Agents move greedily toward the cell with the highest current belief probability, taking one step per timestep along the shortest path. All four protocols share this identical movement policy; we deliberately hold the navigation strategy fixed across conditions so that observed differences in task success and epistemic alignment are attributable solely to communication design, independent of planning quality.

\subsection{Communication Protocols}
\label{sec:protocols}

We evaluate four communication protocols that vary in message content and transmission conditions. The four protocols form a design space along two axes: what information is transmitted (none, point estimate, belief distribution) and when transmission occurs (every timestep, or conditionally on information novelty). Table~\ref{tab:protocols} summarizes their key properties.

\begin{table}[h]
\centering
\caption{Summary of communication protocols.}
\label{tab:protocols}
\begin{tabular}{llll}
\toprule
Protocol & Message Content & Fusion Rule & Gate \\
\midrule
$C_0$ & None & --- & --- \\
$C_1$ & Argmax cell $+$ $H$  & Argmax replacement & None \\
$C_2$ & Top-$k$ cells $+$ probs $+$ $H$ &
        Inv-entropy weighted & None \\
$C_3$ & Top-$k$ cells $+$ probs $+$ $H$  &
        Inv-entropy weighted & $|\Delta H| \geq \theta$ \\
\bottomrule
\end{tabular}
\end{table}

\textbf{$C_0$ --- No Communication.} Agents explore and update beliefs
independently, serving as the baseline against which all communication
protocols are evaluated.

\textbf{$C_1$ --- Semantic Communication.} Each agent broadcasts its
current argmax cell and Shannon entropy $H$ at every timestep. This
protocol represents the natural engineering choice of sharing
conclusions rather than evidence, a point estimate without
uncertainty information. Upon receiving a message, the receiver identifies the sender's reported argmax cell and applies a fixed probability boost to that cell in its own belief, scaled by the sender's reported entropy, lower-entropy senders produce a larger boost.

\textbf{$C_2$ --- Epistemic communication.} Each agent broadcasts a 
compressed summary of its belief distribution: the $k_{\text{msg}} = 5$ 
most probable cells and their associated probabilities, together with 
the sender's Shannon entropy $H$. Transmitting a distribution rather 
than a point estimate preserves the uncertainty structure that the 
fusion rule requires --- a sender with low entropy (concentrated belief) 
contributes more to the receiver's update than one still exploring.

Receivers fuse incoming messages using inverse-entropy weighting:
\begin{equation}
    b'(c) \propto b(c) + \sum_{j \neq i} w_j \cdot m_j(c)
\end{equation}
where $m_j(c)$ is the probability agent $j$ assigns to cell $c$ in its 
transmitted summary, and the fusion weight is:
\begin{equation}
    w_j = \frac{1/H_j}{\sum_{j' \neq i} 1/H_{j'}}, \quad w_j \leq w_{\max} = 0.8
\end{equation}
The cap $w_{\max}$ prevents any single confident sender from 
overwhelming the receiver's belief; when the cap is active the excess weight is redistributed proportionally among the remaining senders. Unlike $C_1$, $C_2$ transmits at every timestep regardless of whether the agent's belief has changed, a design choice that, as we show in Section~\ref{rq2}, drives the malignant epistemic herding dynamic despite the richer message content.

\textbf{$C_3$ --- Gated epistemic communication.} $C_3$ uses identical message content and fusion rules to $C_2$, but conditions each transmission on information novelty: an agent broadcasts only when its Shannon entropy has changed substantially since its last transmission:
\begin{equation}
    |H(t) - H(t_{\text{last}})| \geq \theta, \quad \theta = 0.20 \text{ nats}
\end{equation}
The threshold $\theta$ is selected using the sensitivity analysis reported in Section~\ref{rq4}; we show that the performance is robust in $\theta \in [0.15, 0.35]$.

Entropy serves two complementary roles in this framework. As a 
\textit{transmission gate}, a drop in entropy signals that the agent has acquired new evidence worth sharing, exploration produces gradual entropy reduction, while a target detection causes a sharp drop that immediately triggers a broadcast. As a \textit{fusion weight}, the absolute entropy of the sender determines message influence at the receiver under the inverse-entropy rule: the same discovery event that triggers transmission also maximizes the message's impact. This self-regulating property means $C_3$ transmits precisely when its messages are most informative and most influential, without requiring any explicit coordination of transmission schedules among agents.

\subsection{Network Model}
\label{sec:network}

All communication occurs over a shared logical channel modeled as a 
fully connected broadcast network in which every agent can reach every 
other agent in a single hop. Two sources of network degradation are 
varied independently across experimental conditions.

\textbf{Packet loss} is modeled as congestion-dependent. The probability 
that any message is dropped is a function of the number of concurrent 
transmissions:
\begin{equation}
    P(\text{drop}) = 1 - (1 - p_{\text{base}})^{n}
\end{equation}
where $p_{\text{base}} \in \{0.0, 0.1, 0.3\}$ is the base loss rate 
and $n$ is the number of simultaneous senders. This models a shared 
medium where contention increases with agent activity --- protocols 
that transmit more frequently experience disproportionately higher 
loss rates under this model.

\textbf{Latency} is modeled as a fixed delivery delay of 
$\ell \in \{0, 1, 3\}$ timesteps. A message sent at step $t$ arrives 
at step $t + \ell$. All agents observe and transmit before any fusions are 
applied within a timestep, so beliefs are updated on the previous 
step's messages. This synchronous execution model is standard in 
discrete-time multi-agent simulation.

Packet loss and latency are crossed as independent factors in the 
full factorial design described in Section~\ref{sec:design}. Because 
$C_3$ transmits far fewer messages than $C_1$ or $C_2$, it experiences lower 
effective loss rates under the congestion model.

\section{Metrics and Experimental Design}
\label{sec:metrics}

\subsection{Task Performance Metrics}

Task performance is measured by two metrics. \textbf{Success rate} is
the fraction of episodes in which at least $k$ agents reach the target
cell within the $T = 200$ step time limit, where $k \in \{2, 3, 4\}$
defines the coordination requirement, coordinated arrival ($k=2$), strong coordination
($k=3$), and complete coordination ($k = 4$, all agents must reach the target). \textbf{Time to success} is
the timestep at which the $k$-th agent first reaches the target,
recorded only for successful episodes. Together, these metrics capture
both whether the team succeeded and how efficiently it did so.

\subsection{Epistemic Alignment Metrics}

Beyond task outcomes, we measure the quality of agent belief states
directly using two epistemic alignment metrics recorded at every
timestep throughout each episode.

\textbf{Jensen-Shannon Divergence (JSD)} measures the degree of
disagreement between agent beliefs. At each timestep we compute the
mean pairwise JSD across all agent pairs:

\begin{equation}
    \overline{\text{JSD}} = \frac{1}{\binom{n}{2}}
    \sum_{i < j} \text{JSD}(b_i \| b_j)
\end{equation}

JSD ranges from $0$ to $\log 2$ nats, where $0$ indicates that all
agents hold identical beliefs and $\log 2$ indicates maximum
disagreement. Low JSD signals that agents have converged on a shared
hypothesis --- but as we show, convergence alone does not imply
correctness.

\textbf{Alignment to truth} measures how well the team's beliefs
reflect the actual target location. For each agent $i$, alignment is
the probability mass assigned to the true target cell $c^*$:

\begin{equation}
    A_i = b_i(c^*)
\end{equation}

We report the mean alignment $\bar{A}$ across all agents at each
timestep. A value of $1$ indicates perfect collective knowledge of the
target location; a value near $1/N^2$ indicates beliefs
indistinguishable from the uniform prior. Together, JSD and alignment
characterize the epistemic health of the team: a well-functioning team
exhibits both low JSD (agents agree) and high alignment (they agree on
the right answer).

\subsection{Formalizing Epistemic Alignment}

Let $\mathcal{S}$ denote the set of possible environment states (grid cells), and let each agent $i$ maintain a belief distribution $b_i \in \Delta(\mathcal{S})$, where $\Delta(\mathcal{S})$ is the probability simplex over $\mathcal{S}$. The collective epistemic state of a team of $n$ agents is therefore represented by the set $\{b_1, \dots, b_n\}$.

We define \textit{epistemic alignment} as the degree of similarity among agent belief distributions. Formally, let $D(\cdot \| \cdot)$ denote a divergence measure over probability distributions. We define average pairwise divergence:

\begin{equation}
\mathcal{A}_{div} = \frac{2}{n(n-1)} \sum_{i < j} D(b_i \| b_j)
\end{equation}

In this work, we instantiate $D$ as the Jensen--Shannon divergence (JSD), a symmetric and bounded divergence defined as:

\begin{equation}
JSD(b_i \| b_j) = \frac{1}{2} KL(b_i \| m) + \frac{1}{2} KL(b_j \| m), \quad m = \frac{1}{2}(b_i + b_j)
\end{equation}

JSD is particularly suitable for this setting because it is symmetric, always finite, and bounded in $[0, \log 2]$, making it interpretable as a measure of disagreement between agents.

Epistemic alignment captures only agreement among agents, not correctness. We therefore distinguish between \textit{consensus} (low divergence) and \textit{accuracy} (alignment with ground truth), which we measure separately via posterior mass assigned to the true target location.

This distinction is critical: a system may achieve high epistemic alignment while remaining misaligned with reality, a condition we formalize as MEH in Section \ref{silent_fail4}.

\subsection{Malignant Epistemic Herding as a Collective Failure Mode}
\label{silent_fail4}
Standard task metrics alone cannot distinguish between a team that failed due to insufficient time and one that failed because agents collectively converged on an incorrect hypothesis. We define \textit{malignant epistemic herding} to capture this latter condition.

\begin{equation}
\text{MEH} \iff \mathcal{A}_{div} < \epsilon \ \wedge \ \text{task failed}
\end{equation}

\begin{figure*}[t] 
\centering 
\fbox{\includegraphics[scale=.50]{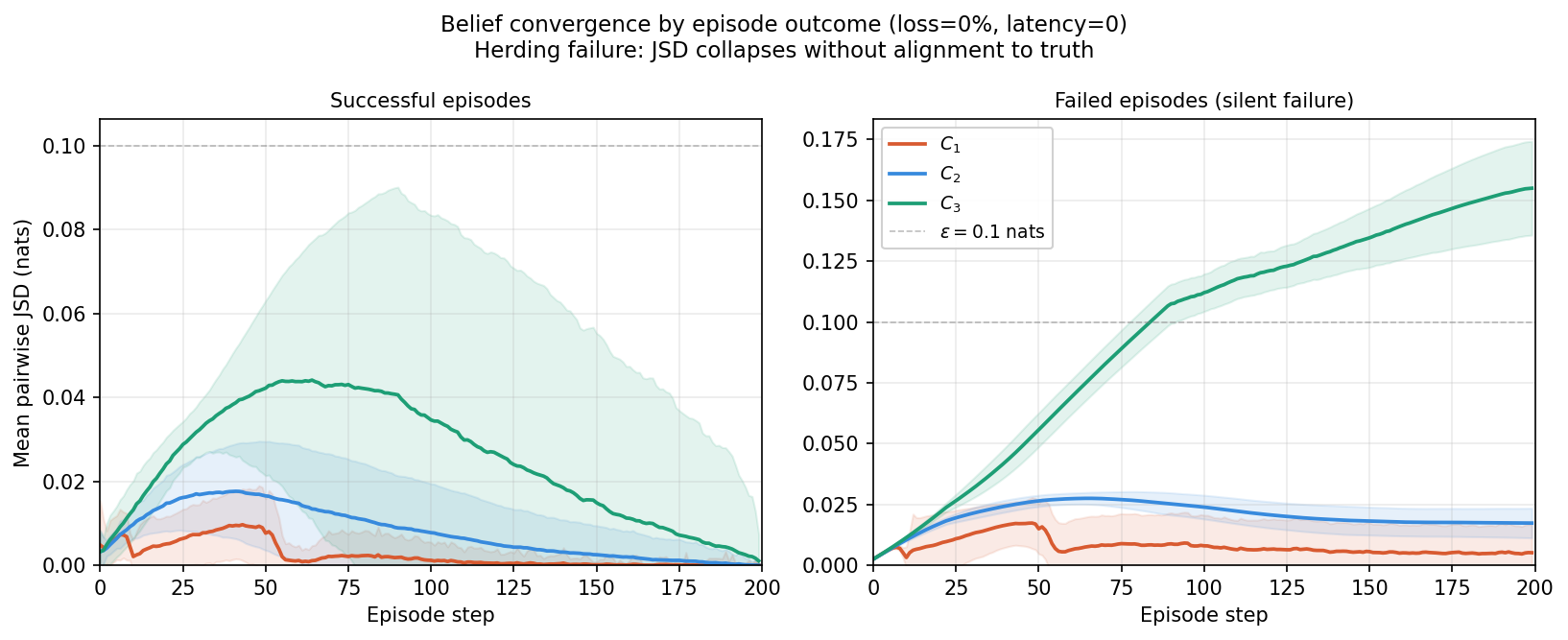}}
\caption{Belief convergence by episode outcome. \textit{Left}: in 
successful episodes, all protocols converge toward zero JSD as agents 
align on the correct hypothesis. \textit{Right}: in failed episodes, 
$C_1$ and $C_2$ collapse well below $\epsilon = 0.1$ nats (dashed), 
indicating coordinated error. $C_3$ failed episodes show increasing 
JSD, indicating independent rather than coordinated failure, a 
qualitatively different and less dangerous failure mode.  Results are shown under zero packet loss and zero latency to isolate communication dynamics from network effects.}
\label{jsd_out} 
\end{figure*}

We set $\epsilon = 0.2$ nats, a conservative threshold corresponding to approximately 14\% of the maximum possible JSD ($\log 2 \approx 0.693$ nats). As we show in Figure \ref{jsd_out}, $C_1$ and $C_2$ routinely reach JSD values below $0.01$ nats, an order of magnitude below this threshold, confirming that the MEH classification is not sensitive to the choice of $\epsilon$.

MEH represents a class of collective failure modes in decentralized systems in which agents achieve strong internal agreement while remaining jointly incorrect. This phenomenon is closely related to information cascades and herding behavior, where early signals disproportionately influence group belief formation, leading to premature convergence on incorrect hypotheses\cite{bikhchandani1992theory}.

In decentralized belief fusion settings, MEH can arise when communication protocols repeatedly reinforce confident but incorrect beliefs. Importantly, MEH is not detectable from agreement alone: low divergence may indicate either successful coordination or coordinated error.

By explicitly distinguishing between epistemic alignment and alignment to ground truth, MEH captures a failure mode that is invisible to traditional coordination metrics but critical in high-stakes applications.

\subsection{Communication Cost Metrics}

To evaluate bandwidth efficiency we record three communication metrics
per episode: \textbf{messages sent} (total transmissions across all
agents), \textbf{bytes transmitted} (total payload volume), and
\textbf{alignment per byte} (final mean alignment divided by bytes
transmitted). The last metric captures the information value delivered
per unit of bandwidth consumed, allowing direct efficiency comparison
between protocols that transmit at different rates and message sizes.
Alignment per byte is undefined for $C_0$, which transmits no
messages, and is therefore excluded from efficiency comparisons.

\subsection{Experimental Design}
\label{sec:design}

We evaluate all four communication protocols under a full factorial
design crossing three additional independent factors:

\begin{itemize}
    \item \textbf{Communication protocol:} $C_0$, $C_1$, $C_2$, $C_3$
    \item \textbf{Packet loss rate:} $p_{\text{base}} \in \{0.0, 0.1, 0.3\}$
    \item \textbf{Latency:} $\ell \in \{0, 1, 3\}$ timesteps
    \item \textbf{Coordination threshold:} $k \in \{2, 3, 4\}$ agents
          required for task success
\end{itemize}

The coordination threshold $k$ is a true experimental factor rather
than a post-hoc filter, as it controls episode termination: an episode
ends as soon as $k$ agents reach the target or the 200-step time limit
is reached. An episode run under $k=2$ terminates as soon as two agents
succeed, producing different belief trajectories and communication
patterns than the same episode run under $k=4$. Each threshold
therefore requires an independent set of simulation runs.

This yields $4 \times 3 \times 3 \times 3 = 108$ conditions in total.
Since $C_0$ is unaffected by network parameters, its conditions serve
as a repeated baseline within each coordination threshold. Each
condition is evaluated over 1,000 independent episodes with
different random seeds controlling target placement
and network behavior. All experiments are fully deterministic given
seed values, ensuring reproducibility. Results are reported as means
with standard deviations across seeds unless otherwise noted.

Each research question maps to a specific subset of conditions. RQ1
compares all four protocols at zero packet loss and zero latency across
all three coordination thresholds, establishing the baseline effect of
communication on task success. RQ2 compares $C_1$ and $C_2$ across
all network conditions using epistemic alignment metrics, with $k=2$
as the primary coordination criterion. RQ3 compares $C_2$ and $C_3$
on communication cost metrics across all network conditions, examining
whether gating preserves performance while reducing bandwidth.

All reported pairwise comparisons between communication protocols on task success rate, final alignment, final JSD, and MEH rate were evaluated using two-sided Mann-Whitney U tests for continuous metrics and two-proportion z-tests for binary rates, with Benjamini-Hochberg correction applied across all tests. All protocol comparisons across primary conditions (zero packet loss, zero latency) are statistically significant at FDR $< 0.05$. Under degraded network conditions, C0 versus C1 success rate comparisons are non-significant at high latency, consistent with C1's known inability to recover from early misdirection under message loss, and C2 versus C3 final alignment comparisons are non-significant at maximum packet loss and latency, where reduced message delivery narrows the behavioral difference between continuous and gated transmission.
\section{Results}

\begin{figure*}[t] 
\centering 
\includegraphics[scale=.75]{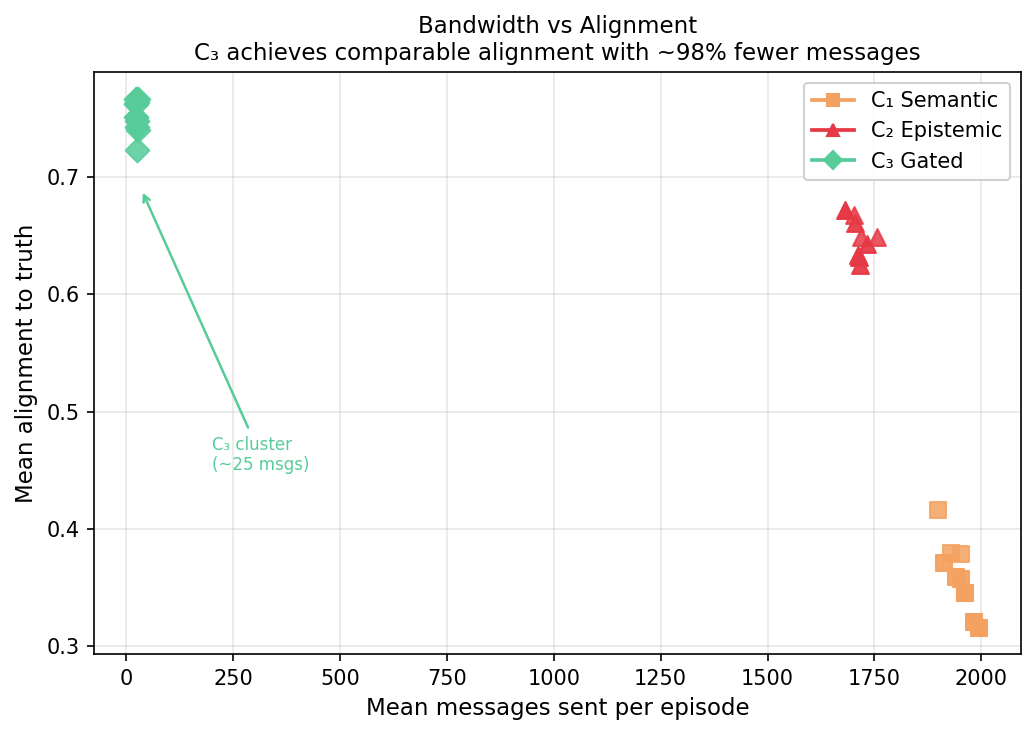} 
\caption{Bandwidth efficiency across communication protocols. Mean alignment 
to truth versus mean messages sent per episode. $C_3$ achieves alignment 
comparable to $C_2$ with over 98\% fewer messages, clustering near the 
origin while $C_1$ and $C_2$ transmit continuously at 1,300--1,600 
messages per episode.}
\label{sys_over} 
\end{figure*}

\begin{figure*}[t] 
\centering 
\fbox{\includegraphics[scale=.55]{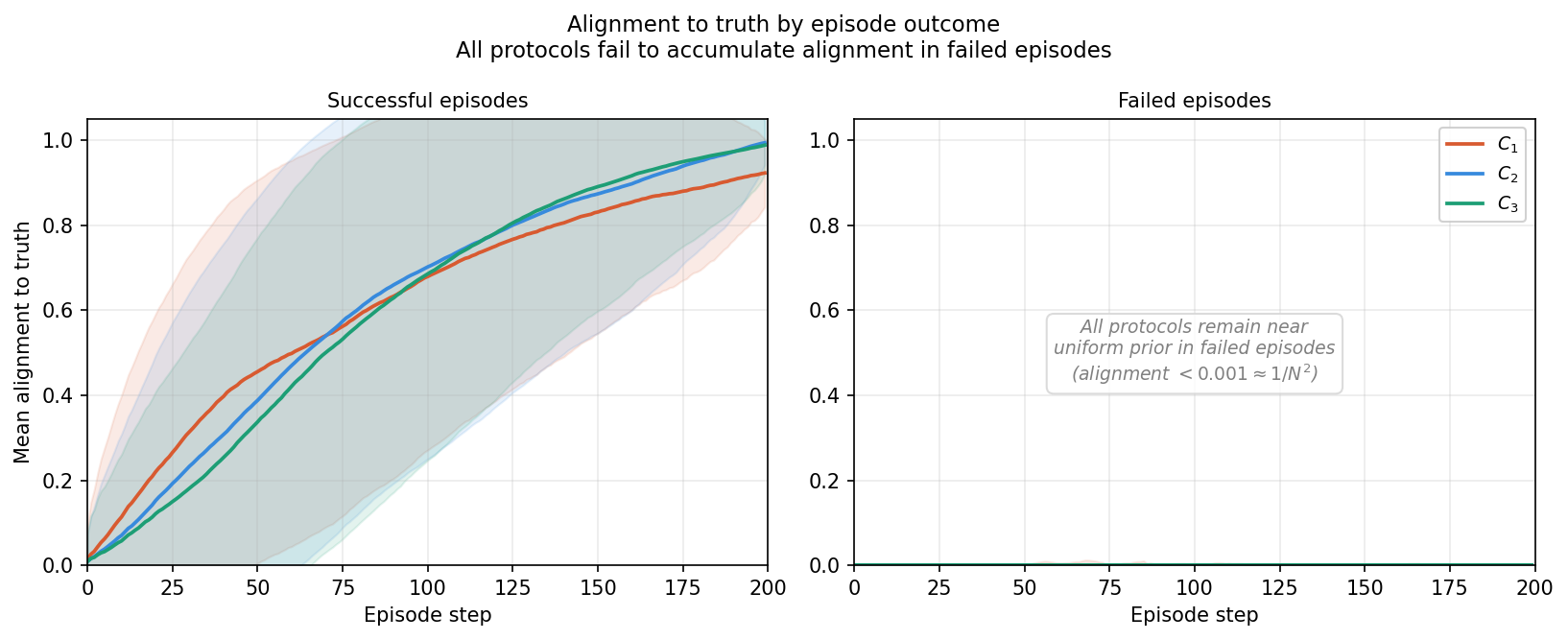}}
\caption{Alignment to truth by episode outcome. \textit{Left}: 
in successful episodes, all protocols accumulate alignment 
monotonically as agents converge on the correct target cell. 
\textit{Right}: in failed episodes, all protocols remain near 
the uniform prior ($\bar{A} < 0.001 \approx 1/N^2$) regardless 
of communication design, confirming that failed episodes involve 
no meaningful belief accumulation toward the true target.}
\label{alignment_out} 
\end{figure*}

\begin{figure}[t] 
\centering 
\fbox{\includegraphics[scale=.65]{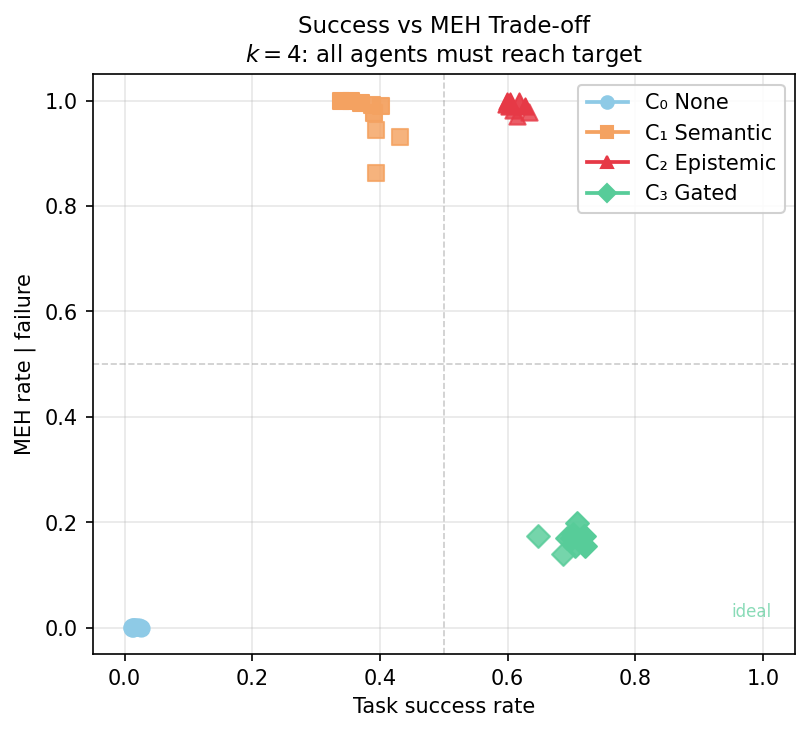}} 
\caption{Task success rate versus MEH rate under the strongest 
coordination requirement ($k = 4$, all agents must reach the target). 
$C_3$ occupies the desirable bottom-right region (high success, low 
MEH). $C_1$ and $C_2$ achieve moderate-to-high success but at 
near-total MEH rates. Marker opacity reflects packet loss rate 
(lighter $=$ higher loss).}
\label{fig:success_vs_silentfail} 
\end{figure}

\subsection{RQ1: How does communication affect coordinated task success?}

Table~\ref{tab:task_performance}  shows task performance across 
communication types and coordination requirements. All three 
communication protocols improve over the no-communication baseline 
($C_0$) in both alignment and success rate. $C_1$ achieves the 
fastest time-to-success among successful episodes ($k{=}2$: 82.1 
vs.\ 86.4 for $C_2$), but its overall success rate is substantially 
lower (0.388 vs.\ 0.639). $C_3$ achieves the highest alignment and 
success rates across all coordination thresholds, at a modest cost 
in time-to-success relative to $C_2$, a tradeoff we examine in 
RQ3.

\begin{table}[t]
\caption{Task performance by communication type and coordination requirement: mean alignment to truth ($\uparrow$), task success rate ($\uparrow$), and time-to-success in timesteps ($\downarrow$). Bold indicates best per column.}
\label{tab:task_performance}
\begin{tabular}{lrrrrrrrrr}
\toprule
 & \multicolumn{3}{c}{Align $\uparrow$} & \multicolumn{3}{c}{Success $\uparrow$} & \multicolumn{3}{c}{Time $\downarrow$} \\
\cmidrule(lr){2-4} \cmidrule(lr){5-7} \cmidrule(lr){8-10}
Comm & $k{=}2$ & $k{=}3$ & $k{=}4$ & $k{=}2$ & $k{=}3$ & $k{=}4$ & $k{=}2$ & $k{=}3$ & $k{=}4$ \\
\midrule
$C_0$ & 0.303 & 0.329 & 0.336 & 0.404 & 0.117 & 0.016 & 121.3 & 147.4 & 162.6 \\
$C_1$ & 0.345 & 0.358 & 0.360 & 0.388 & 0.388 & 0.385 &  \textbf{82.1} &  \textbf{89.7} & \textbf{100.8} \\
$C_2$ & 0.633 & 0.653 & 0.648 & 0.639 & 0.628 & 0.611 &  86.4 &  95.7 & 105.7 \\
$C_3$ & \textbf{0.700} & \textbf{0.740} & \textbf{0.752} & \textbf{0.734} & \textbf{0.723} & \textbf{0.698} &  93.7 & 103.7 & 115.1 \\
\bottomrule
\end{tabular}
\end{table}

\subsection{RQ2: Does epistemic communication prevent malignant epistemic herding?}
\label{rq2}
\begin{table}[t]
\caption{Communication efficiency by communication type and coordination requirement: alignment per byte (APB, $\times 10^{-6}$, $\uparrow$), message count ($\downarrow$), and MEH rate ($\downarrow$). $C_0$ is not included since no communication takes place. Bold indicates best per column.}
\label{tab:efficiency}
\begin{tabular}{lrrrrrrrrr}
\toprule
 & \multicolumn{3}{c}{APB $\uparrow$} & \multicolumn{3}{c}{Msgs $\downarrow$} & \multicolumn{3}{c}{MEH $\downarrow$} \\
\cmidrule(lr){2-4} \cmidrule(lr){5-7} \cmidrule(lr){8-10}
Comm & $k{=}2$ & $k{=}3$ & $k{=}4$ & $k{=}2$ & $k{=}3$ & $k{=}4$ & $k{=}2$ & $k{=}3$ & $k{=}4$ \\
\midrule
$C_1$ &    26.6 &    27.1 &    26.5 & 1856 & 1891 & 1947 & 0.963 & 0.969 & 0.966 \\
$C_2$ &    11.8 &    11.5 &    10.8 & 1537 & 1622 & 1715 & 0.990 & 0.987 & 0.989 \\
$C_3$ & \textbf{1091.4} & \textbf{887.5} & \textbf{838.9} & \textbf{19} & \textbf{24} & \textbf{26} & \textbf{0.062} & \textbf{0.093} & \textbf{0.168} \\
\bottomrule
\end{tabular}
\end{table}

Table~\ref{tab:efficiency} shows that epistemic encoding alone does not 
prevent MEH. $C_2$'s MEH rate (0.990 at $k{=}2$) 
is marginally \emph{higher} than $C_1$'s (0.963), despite transmitting 
richer belief information. We attribute this to a self-reinforcing 
convergence dynamic under high message frequency: once any agent develops 
a confident but incorrect belief, it broadcasts at high inverse-entropy 
weight, nudging peers toward its hypothesis, reducing their entropy, and 
further amplifying its influence in subsequent steps. This feedback loop 
drives rapid belief convergence regardless of message content, locking 
the team onto an incorrect hypothesis with no visible sign of failure. 
High-frequency epistemic communication therefore improves task success 
while paradoxically worsening epistemic health.

\subsection{RQ3: Does entropy-gated communication resolve the frequency-failure tradeoff?}

Entropy-delta gating breaks the feedback loop identified in RQ2. $C_3$ 
transmits only when an agent's belief has shifted significantly 
($|\Delta H| \geq \theta$), reducing message volume from approximately 
1,500-1,900 to fewer than 30 per episode, a reduction of over 98\%. This 
sparse messaging pattern prevents the continuous reinforcement that drives 
MEH in $C_2$, yielding a MEH rate of 0.062 at 
$k{=}2$ compared to 0.990 for $C_2$. The mean alignment-per-byte efficiency of $C_3$ across coordination thresholds ($939 \times 10^{-6}$) is over $80\times$ that of $C_2$ 
($11.4 \times 10^{-6}$), confirming that sparse epistemic messages 
deliver substantially more information value per unit of bandwidth than 
continuous transmission.

\subsection{RQ4: Sensitivity to Entropy Threshold}
\label{rq4}

\begin{figure*}[t] 
\centering 
\fbox{\includegraphics[scale=.65]{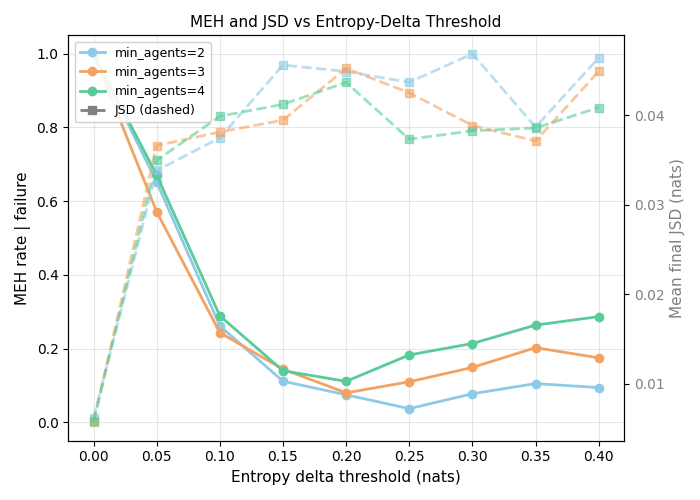}} 
\caption{MEH and JSD plotted against entropy delta threshold varying between 0 (full communication) and 0.40 (significantly reduced communication).}
\label{delta_sweep} 
\end{figure*}

We evaluate the sensitivity of entropy-delta gating ($C_3$) to the threshold parameter $\theta$, which controls when agents transmit messages. While previous results fix $\theta = 0.20$, the choice of threshold directly governs the tradeoff between communication frequency and responsiveness to new information.

We vary $\theta$  in 0.05 increments ranging from $[0.00, 0.40]$ and measure task success, MEH rate, and total message count. We test each delta-gate with 1,000 iterations and requiring three agents to successfully coordinate in identifying the target for task success.

Across $\theta \in [0.20, 0.35]$, message volume varies by fewer than 
three messages per episode on average, a difference that is operationally 
negligible relative to the $T = 200$ step budget. Within the bandwidth plateau, $\theta = 0.20$ has the lowest MEH rate (0.089). We therefore select $\theta = 0.20$ as the fixed entropy-gate value for other experiments to minimize coordinated error while maintaining bandwidth efficiency. These results are shown in Table \ref{tab:delta_sensitivity}. Additionally, the associated MEH and pairwise JSD values for $\theta \in [0.20, 0.35]$ across all agent coordination levels are shown in Figure \ref{delta_sweep}. 

\begin{table}[t]
\centering
\caption{Sensitivity of entropy-delta gating ($C_3$) to threshold $\theta$
         ($k=3$, 1{,}000 episodes). $\theta = 0$ corresponds to ungated
         epistemic communication (equivalent to $C_2$). Message volume
         is approximately constant across $\theta \in [0.20, 0.35]$
         ($\approx 21$--$23$ messages per episode), while MEH
         reaches a minimum at $\theta = 0.20$. We therefore select
         $\theta = 0.20$ as the operating point that minimizes coordinated
         error within the low-bandwidth plateau. Bold indicates the best
         value per column; \underline{underline} marks the selected
         operating point.}
\label{tab:delta_sensitivity}
\begin{tabular}{rrrrrrrr}
\toprule
$\theta$ & Success\,$\uparrow$ & Time\,$\downarrow$ & Silent Fail\,$\downarrow$ & JSD$^\dagger$ & Align\,$\uparrow$ & Msgs\,$\downarrow$ & APB\,$\uparrow$ \\
\midrule
0.00 & 0.629 & \textbf{93.7} & 1.000 & 0.006 & 0.657 & 1604.6 &  11.7 \\
0.05 & 0.615 & 96.9 & 0.631 & 0.035 & 0.636 &   51.3 & 355.2 \\
0.10 & 0.653 & 96.3 & 0.265 & 0.039 & 0.670 &   32.4 & 592.7 \\
0.15 & 0.658 & 98.9 & 0.132 & 0.042 & 0.680 &   25.8 & 754.0 \\
\underline{0.20} & \underline{0.695} & \underline{101.5} & \underline{\textbf{0.089}} & \underline{0.045} & \underline{0.711} & \underline{22.9} & \underline{893.5} \\
0.25 & 0.720 & 102.8 & 0.110 & 0.041 & 0.740 &   22.4 & 951.4 \\
0.30 & 0.726 & 102.2 & 0.147 & 0.041 & 0.753 &   21.3 & 1018.9 \\
0.35 & \textbf{0.755} & 104.4 & 0.190 & 0.038 & \textbf{0.790} &   20.8 & 1095.8 \\
0.40 & 0.738 & 104.1 & 0.185 & \textbf{0.044} & 0.776 & \textbf{19.2} & \textbf{1164.7} \\
\bottomrule
\end{tabular}
\end{table}
\footnote{$^\dagger$ JSD has no universally preferred direction in this study. 
Low JSD in C$_1$ and C$_2$ reflects malignant epistemic herding; 
moderate JSD in C$_3$ reflects maintained epistemic diversity. 
See Section~4.3.}

\subsection{Scaling Analysis}

\begin{figure}[htbp]
\centering
\fbox{\includegraphics[width=\textwidth]{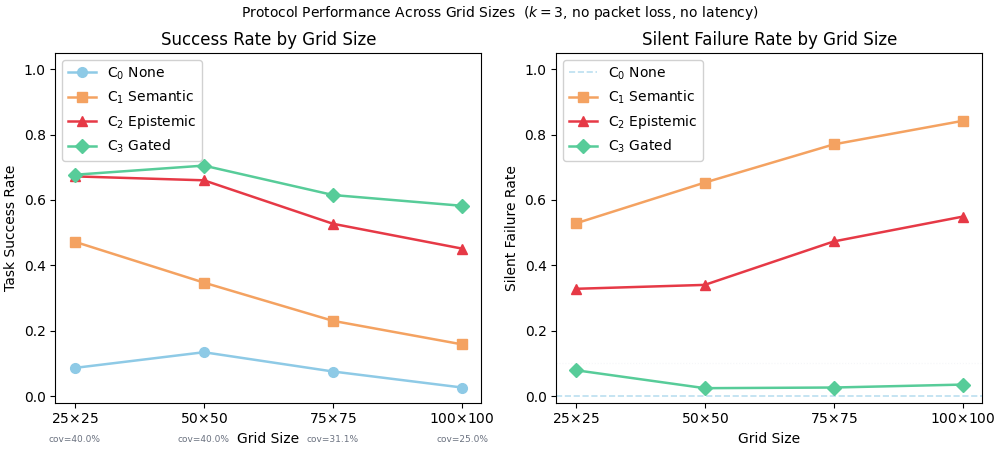}}
\caption{Protocol performance across grid sizes ($k=3$, no packet loss, 
no latency). \textit{Left}: task success rate decreases with grid size 
for all protocols, but C$_3$ maintains the largest absolute advantage 
over C$_2$ at every scale. \textit{Right}: MEH rate increases 
monotonically with grid size for C$_1$ and C$_2$, while C$_3$ remains 
near zero across all conditions. C$_0$ is shown as a dashed reference 
line at zero MEH, as independent agents cannot exhibit 
communication-driven belief convergence. Per-agent coverage ratios 
are annotated below the x-axis of the left panel.}
\label{fig:scaling}
\end{figure}

While the predominant results are measured on a $50 \times 50$ grid, 
we evaluate generalizability across state space sizes by running 1,000 
episodes for $k=3$ on grids of $25 \times 25$, $50 \times 50$, 
$75 \times 75$, and $100 \times 100$ with episode budgets of 
$T \in \{50, 200, 350, 500\}$ respectively (Table~\ref{scaling}). 
These budgets yield maximum theoretical per-agent coverage ratios of 
$\{40.0\%, 40.0\%, 31.1\%, 25.0\%\}$ for grid sizes 
$N \in \{25, 50, 75, 100\}$. The two smaller grids are 
coverage-matched; the two larger grids operate under progressively 
tighter budgets, making the larger-scale conditions more demanding 
and providing a conservative test of protocol robustness.

The qualitative ordering of protocols is consistent across all grid 
sizes: C$_3$ achieves the highest success rate and alignment at every 
scale, C$_2$ ranks second on both metrics, and MEH rates 
follow the inverse pattern. Notably, the MEH gap between 
C$_2$ and C$_3$ widens systematically with grid size --- from 0.249 
at $25 \times 25$ to 0.514 at $100 \times 100$ --- consistent with 
the theoretical prediction that the penalty for premature belief 
convergence grows with state space size. C$_1$'s MEH rate 
increases monotonically from 0.528 to 0.842 as the grid grows, 
suggesting that point-estimate communication becomes increasingly 
hazardous as search complexity increases and the probability of early 
misdirection compounds over a larger state space. These trends are highlighted in Figure \ref{fig:scaling}.

JSD values are approximately stable across grid sizes for all 
communicating protocols, but through distinct mechanisms. C$_1$ and 
C$_2$ maintain consistently low JSD (0.003--0.011) because 
high-frequency communication drives belief convergence within the 
first 50--75 timesteps regardless of state space size, after which 
agents share a common hypothesis whether correct or not. C$_3$ 
maintains moderate and stable JSD (0.042--0.049) through the 
entropy-delta gate, which actively preserves epistemic diversity by 
suppressing redundant transmissions, a scale-invariant property 
that reflects designed equilibrium between convergence and diversity 
rather than convergence alone. C$_0$'s decreasing JSD with grid size 
reflects a measurement artifact: as the grid grows relative to the 
fixed episode budget, agents' beliefs remain near-uniform over the 
majority of unvisited cells, producing apparent agreement in pairwise 
JSD despite increasingly isolated local observations with no 
overlapping coverage. This suggests JSD may understate epistemic 
divergence in large sparse belief settings, and motivates future work 
on coverage-adjusted divergence metrics that account for the 
proportion of the state space each agent has observed.

\begin{table}[htbp]
\centering
\caption{Scaling Analysis Across Grid Sizes ($k=3$, no packet loss, no latency). Each condition uses 1{,}000 episodes (500 for $100 \times 100$). Bold indicates best value per metric within each grid size block.}
\label{scaling}
\small
\begin{tabular}{llrrrrrr}
\toprule
Protocol & Grid & $T$ & Success $\uparrow$ & MEH Rate $\downarrow$ & Align $\uparrow$ & JSD \\
\midrule
\multicolumn{7}{l}{\textit{Grid: $25\times 25$, $T=50$}} \\
\addlinespace[2pt]
C0 & $25\times25$ & 50 & 0.086 & \textbf{0.000} & 0.326 & 0.4122 \\
C1 & $25\times25$ & 50 & 0.472 & 0.528 & 0.473 & \textbf{0.0038} \\
C2 & $25\times25$ & 50 & 0.672 & 0.328 & \textbf{0.750} & 0.0108 \\
C3 & $25\times25$ & 50 & \textbf{0.677} & 0.079 & 0.745 & 0.0423 \\
\midrule
\multicolumn{7}{l}{\textit{Grid: $50\times 50$, $T=200$}} \\
\addlinespace[2pt]
C0 & $50\times50$ & 200 & 0.134 & \textbf{0.000} & 0.341 & 0.4088 \\
C1 & $50\times50$ & 200 & 0.347 & 0.653 & 0.319 & \textbf{0.0030} \\
C2 & $50\times50$ & 200 & 0.660 & 0.340 & 0.691 & 0.0052 \\
C3 & $50\times50$ & 200 & \textbf{0.705} & 0.024 & \textbf{0.723} & 0.0431 \\
\midrule
\multicolumn{7}{l}{\textit{Grid: $75\times 75$, $T=350$}} \\
\addlinespace[2pt]
C0 & $75\times75$ & 350 & 0.075 & \textbf{0.000} & 0.273 & 0.3723 \\
C1 & $75\times75$ & 350 & 0.230 & 0.770 & 0.211 & \textbf{0.0036} \\
C2 & $75\times75$ & 350 & 0.527 & 0.473 & 0.541 & 0.0054 \\
C3 & $75\times75$ & 350 & \textbf{0.615} & 0.026 & \textbf{0.634} & 0.0491 \\
\midrule
\multicolumn{7}{l}{\textit{Grid: $100\times 100$, $T=500$}} \\
\addlinespace[2pt]
C0 & $100\times100$ & 500 & 0.026 & \textbf{0.000} & 0.210 & 0.3269 \\
C1 & $100\times100$ & 500 & 0.158 & 0.842 & 0.145 & \textbf{0.0033} \\
C2 & $100\times100$ & 500 & 0.451 & 0.549 & 0.472 & 0.0049 \\
C3 & $100\times100$ & 500 & \textbf{0.582} & 0.035 & \textbf{0.608} & 0.0478 \\
\bottomrule
\end{tabular}
\end{table}

\section{Discussion}

\subsection{Message Content and Frequency Govern Distinct Outcomes}
Our results demonstrate that communication design in multi-agent search involves two separable concerns that prior work has not clearly distinguished. Message content, whether an agent shares a point estimate or a compressed belief distribution, determines how well the team coordinates on the correct target when coordination occurs. Message frequency determines whether agents preserve the belief diversity necessary to keep searching when early evidence is misleading. These two properties can work against each other: $C_1$ converges faster than $C_2$ in successful episodes (82.1 vs.\ 86.4 timesteps at $k{=}2$), but its overall success rate is 
substantially lower (0.388 vs.\ 0.639) because point-estimate messaging provides insufficient information for agents to recover from early misdirection. $C_2$ improves success by transmitting richer belief content, but this same richness accelerates convergence when messages arrive at every timestep, trading speed for robustness. The practical implication is that optimizing message content alone is insufficient, agentic communication protocol design for search and coordination tasks must consider both what is transmitted and how often.

C$_1$ and C$_2$ differ in both message content and payload size,
C$_1$ transmits 7 bytes per message while C$_2$ transmits 35 bytes. 
This means their performance differences reflect a combination of 
information richness and bandwidth usage. A fully controlled comparison 
would require equalizing bytes transmitted across protocols. We note 
however that the alignment-per-byte metric directly accounts for this 
by normalizing alignment gains against transmission cost, and 
C$_3$ dominates both C$_1$ and C$_2$ on this metric by over 
$80\times$ regardless of their byte-level differences (Table~4).

\subsection{The MEH Mechanism}
MEH arises from a self-reinforcing dynamic inherent to inverse-entropy fusion under high message frequency. Agents broadcast their beliefs with high frequency, causing a gradual collapse of beliefs toward a single shared hypothesis. This methodology works when agents have a correct hypothesis of the world, but fails emphatically when agents converge to the wrong belief map. The result in erroneous beliefs is a rapid convergence indistinguishable from successful coordination in standard task metrics. This mechanism is not unique to our fusion rule: any weighted fusion scheme that privileges confident senders will exhibit analogous dynamics when messages are frequent. Prior work on distributed Bayesian fusion, including Covariance 
Intersection and its variants, has focused on preventing overconfidence in individual belief updates. Our findings suggest this is insufficient, team-level belief diversity requires active protection even when individual updates are well-calibrated.

\subsection{Entropy Gating as a Coordination Primitive}
Entropy-delta gating resolves the content-frequency tension by making transmission dependent on information novelty. An agent that is still exploring transmits infrequently, preserving its independence and contributing to search coverage. An agent that has found strong evidence transmits immediately with an associated high fusion weight, causing rapid convergence to the correct hypothesis. The dual use of entropy as a transmission gate and fusion weight means that the protocol is self-regulating, where the conditions that trigger a message are the same conditions that cause a message to be most influential. To our knowledge, this complementary use of entropy has not been previously characterized in the event-triggered communication literature, which primarily focuses on state deviation triggers as opposed to information-gain methods.

\subsection{Generalization Beyond Grid-Based Search}

While our experiments are conducted in a discrete grid environment with a single target, the phenomena observed here are not specific to this domain. Any decentralized system in which agents maintain probabilistic beliefs and exchange information under uncertainty may exhibit similar dynamics.

Examples include multi-robot exploration, distributed sensor fusion, swarm coordination, and networked AI agents operating over shared information environments. In such systems, communication protocols that repeatedly reinforce confident estimates may induce premature belief convergence, leading to MEH.

The entropy-gating mechanism proposed here is agnostic to the underlying state space and can be applied to continuous or structured belief representations, provided a suitable uncertainty measure is available. Future work should examine these dynamics in more complex environments, including adversarial settings and learned communication protocols.

\subsection{Limitations}
\label{sec:limitations}

The findings reported here should be interpreted in light of several simplifying assumptions that bound the scope of our conclusions.

\textbf{Movement and planning.} Agents follow a greedy policy toward their current belief argmax, which is deliberately simple but not optimal. Information-theoretic planners that maximize expected entropy reduction~\cite{charrow2015information, frew2009information} would likely improve task 
success across all protocols by reducing redundant coverage. Our choice to hold movement fixed isolates communication as the experimental variable, but means our results characterize protocol differences under suboptimal navigation. Whether the content-frequency tradeoff and malignant epistemic herding dynamics persist under more sophisticated planners is an open question. The scaling analysis in Section~5.5 shows that C$_3$'s advantages over C$_2$ in MEH rate and alignment persist and widen across grid sizes from $25 \times 25$ to $100 \times 100$, suggesting the content-frequency tradeoff is not an artifact of the movement policy under the specific grid size used in the primary experiment.

\textbf{Agent homogeneity.} All agents are identical in sensor capability, belief representation, and movement policy. Real deployments typically involve heterogeneous teams where agents differ in sensing range, computational capacity, or communication reach. 
Heterogeneity would complicate the inverse-entropy fusion rule, where agents with systematically better sensors would accumulate lower entropy faster, potentially dominating belief fusion regardless of evidence quality. Whether asymmetric epistemic authority requires 
modified weighting schemes is a natural extension of this work.

\textbf{Environment structure.} Our experiments use a single stationary target in a bounded discrete grid. Multi-target environments would introduce belief partitioning challenges where 
agents may converge on different targets, producing high JSD that is not indicative of poor coordination. Dynamic targets would require belief decay mechanisms not present in our 
current update rule. The discrete grid also precludes continuous state spaces where entropy computation and top-$k$ summarization would require different approximations.

\textbf{Sensor model.} We assume perfect binary detection within the field of view, the target is observed with certainty when within range. Introducing false negatives and false positives 
would add a confound between sensor quality and communication quality, which we deliberately avoided to isolate the effect of protocol design. The interaction between noisy sensing and 
epistemic communication under our fusion rule is a meaningful direction for future work.

\textbf{Network model.} Our congestion-dependent packet loss model assumes a shared broadcast medium with no adversarial behavior. Byzantine agents that inject false beliefs, jamming 
that selectively targets high-confidence transmissions, or asymmetric network topologies where not all agents can reach all others would each stress the protocol design in ways not 
captured here. The MEH mechanism may be accelerated or disrupted by adversarial communication, depending on whether the adversary targets belief content or transmission timing.

\textbf{Synchronous execution.} All agents observe, transmit, and fuse within synchronized timesteps. Real distributed systems are asynchronous. Agents act at different rates and messages arrive out of order. While $C_3$'s sparse messaging may confer robustness advantages under asynchrony for the same reasons it performs well under latency, this has not been formally characterized.

\section{Conclusion}
We investigate how communication protocol design affects both task performance and epistemic diversity in decentralized multi-agent search. Across a full factorial simulation design varying communication type, packet loss, latency, and coordination requirement, we found that both message content and frequency cause downstream ramifications in both task success and epistemic diversity.

Semantic communication ($C_1$) enables fast convergence in successful episodes but achieves low overall success rates and near-total MEH, as point estimates provide insufficient information for agents to recover from poor early episode predictions. Epistemic communication ($C_2$) substantially improves task success by transmitting richer belief content and converges to successful task completion faster than entropy-delta gating ($C_3$). However, this early convergence comes at a cost where the high-frequency message fusion also drives rapid belief convergence on incorrect hypotheses, producing MEH. Entropy-delta gating resolves this tension by making transmission contingent on information novelty. While slower convergence to truth in successful search scenarios, it reduces message volume by ~98\%, achieves the highest alignment to truth, success rate, and epistemic diversity during search across all coordination thresholds.

Our central contribution is the identification and formalization of MEH as a team-level phenomenon distinct from individual overconfidence, and the demonstration that it can be substantially mitigated through adaptive gating without sacrificing coordination quality. We further show that entropy serves a dual purpose in this framework as both a transmission gate and a fusion weight, making entropy-delta gating a lightweight and self-regulating coordination primitive.

\subsection{Future Work}
\label{future}
This work can be extended in several directions. The greedy movement policy could be replaced with information-theoretic planners that maximize expected belief entropy reduction, allowing optimization of both communication and movement. Heterogeneous agent teams that vary in sensor quality, capacity, or communication range could test whether inverse-entropy weighting remains effective when agents have asymmetric epistemic authority. Additionally, multi-target and dynamic-target environments would introduce belief partitioning challenges not present in our single-target design. Finally, learned communication protocols trained via multi-agent reinforcement learning could be evaluated against our hand-designed protocols as an upper bound on achievable coordination efficiency.

\backmatter

\section*{Declarations}

\subsection*{Funding}
Not applicable.

\subsection*{Conflict of interest}
The authors declare no competing interests.

\subsection*{Ethics approval}
Not applicable.

\subsection*{Consent for publication}
Not applicable.

\subsection*{Data availability}
Not applicable. All results are reproducible from the 
code repository.

\subsection*{Code availability}
All simulation code is publicly available at 
\url{https://github.com/davidthfarr/agent2agent}. 
Claude Code (Anthropic) was used to assist in 
simulation code development.

\subsection*{Author contributions}
\textbf{David Farr}: Conceptualization, Methodology, 
Software, Formal analysis, Investigation, 
Data curation, Writing --- original draft, 
Writing --- review and editing, Visualization.
\textbf{Iain Cruickshank}: Conceptualization, 
Writing --- review and editing, Supervision.
\textbf{Kate Starbird}: Writing --- review and editing, 
Supervision.
\textbf{Jevin West}: Writing --- review and editing, 
Supervision, Funding acquisition.

\bibliography{sn-bibliography}

\bigskip
\newpage
\begin{appendices}

\appendix
\section{Full Experimental Results}
Tables A1-A3 report means across 1,000 episodes per condition for all 108 experimental conditions (36 per coordination threshold $k \in \{2, 3, 4\}$). Network conditions vary packet loss rate $p_{\text{base}} \in \{0, 10, 30\}\%$ and latency $\ell \in \{0, 1, 3\}$ timesteps. All pairwise protocol comparisons are statistically significant at FDR $< 0.05$ after Benjamini-Hochberg correction unless noted in Section~4.6.

\begin{table}[htbp]
\centering
\caption{Task Performance, $k=2$ agents required. Values are means across 1,000 episodes. TTS = time to success (steps). Bold indicates best value per metric.}
\label{tab:appendix_task_a}
\small
\begin{tabular}{llrrrrrr}
\toprule
Protocol & Loss (\%) & Latency & Success $\uparrow$ & TTS Mean $\downarrow$ & TTS SD \\
\midrule
C0 & 0 & 0 & 0.370 & 123.7 & 47.4 \\
C0 & 0 & 1 & 0.388 & 121.6 & 47.5 \\
C0 & 0 & 3 & 0.397 & 122.6 & 45.5 \\
C0 & 10 & 0 & 0.372 & 117.6 & 48.3 \\
C0 & 10 & 1 & 0.396 & 125.1 & 46.8 \\
C0 & 10 & 3 & 0.397 & 121.2 & 49.2 \\
C0 & 30 & 0 & 0.407 & 126.3 & 46.2 \\
C0 & 30 & 1 & 0.380 & 119.8 & 48.7 \\
C0 & 30 & 3 & 0.401 & 121.7 & 49.4 \\
\addlinespace[4pt]
C1 & 0 & 0 & 0.361 & 82.4 & 42.5 \\
C1 & 0 & 1 & 0.361 & 79.0 & 43.2 \\
C1 & 0 & 3 & 0.351 & 79.7 & 41.8 \\
C1 & 10 & 0 & 0.427 & 79.7 & 44.2 \\
C1 & 10 & 1 & 0.391 & \textbf{78.5} & 42.8 \\
C1 & 10 & 3 & 0.372 & 81.1 & 44.9 \\
C1 & 30 & 0 & 0.399 & 83.2 & 46.3 \\
C1 & 30 & 1 & 0.427 & 86.1 & 47.8 \\
C1 & 30 & 3 & 0.391 & 85.0 & 48.2 \\
\addlinespace[4pt]
C2 & 0 & 0 & 0.642 & 83.4 & 43.6 \\
C2 & 0 & 1 & 0.638 & 85.2 & 44.6 \\
C2 & 0 & 3 & 0.614 & 88.7 & 46.5 \\
C2 & 10 & 0 & 0.649 & 83.3 & 44.1 \\
C2 & 10 & 1 & 0.616 & 81.9 & 43.1 \\
C2 & 10 & 3 & 0.634 & 89.9 & 44.9 \\
C2 & 30 & 0 & 0.645 & 86.9 & 43.2 \\
C2 & 30 & 1 & 0.606 & 85.2 & 42.9 \\
C2 & 30 & 3 & 0.615 & 86.5 & 41.3 \\
\addlinespace[4pt]
C3 & 0 & 0 & 0.713 & 92.6 & 44.3 \\
C3 & 0 & 1 & 0.722 & 91.0 & 43.1 \\
C3 & 0 & 3 & 0.693 & 89.6 & 40.3 \\
C3 & 10 & 0 & 0.704 & 89.4 & 41.2 \\
C3 & 10 & 1 & 0.712 & 89.9 & 42.7 \\
C3 & 10 & 3 & 0.721 & 94.1 & 45.1 \\
C3 & 30 & 0 & 0.710 & 91.2 & 43.0 \\
C3 & 30 & 1 & 0.714 & 93.7 & 44.2 \\
C3 & 30 & 3 & \textbf{0.730} & 95.5 & 43.9 \\
\bottomrule
\end{tabular}
\end{table}

\begin{table}[htbp]
\centering
\caption{Task Performance, $k=3$ agents required. Values are means across 1,000 episodes. TTS = time to success (steps). Bold indicates best value per metric.}
\label{tab:appendix_task_b}
\small
\begin{tabular}{llrrrrrr}
\toprule
Protocol & Loss (\%) & Latency & Success $\uparrow$ & TTS Mean $\downarrow$ & TTS SD \\
\midrule
C0 & 0 & 0 & 0.128 & 148.1 & 40.1 \\
C0 & 0 & 1 & 0.112 & 148.8 & 36.9 \\
C0 & 0 & 3 & 0.118 & 148.7 & 35.8 \\
C0 & 10 & 0 & 0.113 & 151.5 & 38.8 \\
C0 & 10 & 1 & 0.115 & 148.7 & 37.1 \\
C0 & 10 & 3 & 0.120 & 139.2 & 38.1 \\
C0 & 30 & 0 & 0.104 & 150.0 & 34.8 \\
C0 & 30 & 1 & 0.125 & 151.9 & 35.7 \\
C0 & 30 & 3 & 0.118 & 143.1 & 37.2 \\
\addlinespace[4pt]
C1 & 0 & 0 & 0.339 & 86.7 & 36.9 \\
C1 & 0 & 1 & 0.382 & 89.4 & 40.0 \\
C1 & 0 & 3 & 0.353 & 85.8 & 36.3 \\
C1 & 10 & 0 & 0.399 & \textbf{85.3} & 35.7 \\
C1 & 10 & 1 & 0.375 & 88.0 & 37.6 \\
C1 & 10 & 3 & 0.382 & 91.9 & 40.3 \\
C1 & 30 & 0 & 0.386 & 88.3 & 39.8 \\
C1 & 30 & 1 & 0.404 & 91.7 & 40.6 \\
C1 & 30 & 3 & 0.379 & 91.9 & 39.0 \\
\addlinespace[4pt]
C2 & 0 & 0 & 0.629 & 94.0 & 39.6 \\
C2 & 0 & 1 & 0.621 & 93.9 & 40.9 \\
C2 & 0 & 3 & 0.591 & 95.3 & 38.8 \\
C2 & 10 & 0 & 0.635 & 92.1 & 37.1 \\
C2 & 10 & 1 & 0.636 & 93.7 & 39.2 \\
C2 & 10 & 3 & 0.624 & 95.9 & 37.8 \\
C2 & 30 & 0 & 0.642 & 97.1 & 39.3 \\
C2 & 30 & 1 & 0.615 & 99.5 & 40.6 \\
C2 & 30 & 3 & 0.623 & 99.3 & 37.8 \\
\addlinespace[4pt]
C3 & 0 & 0 & 0.698 & 99.9 & 38.5 \\
C3 & 0 & 1 & 0.704 & 99.4 & 38.9 \\
C3 & 0 & 3 & 0.702 & 102.2 & 38.0 \\
C3 & 10 & 0 & 0.700 & 101.4 & 39.5 \\
C3 & 10 & 1 & 0.703 & 100.6 & 39.6 \\
C3 & 10 & 3 & \textbf{0.706} & 103.8 & 39.2 \\
C3 & 30 & 0 & 0.688 & 101.7 & 39.0 \\
C3 & 30 & 1 & 0.689 & 104.9 & 40.2 \\
C3 & 30 & 3 & 0.689 & 104.7 & 39.3 \\
\bottomrule
\end{tabular}
\end{table}

\begin{table}[htbp]
\centering
\caption{Task Performance, $k=4$ agents required. Values are means across 1,000 episodes. TTS = time to success (steps). Bold indicates best value per metric.}
\label{tab:appendix_task_c}
\small
\begin{tabular}{llrrrrrr}
\toprule
Protocol & Loss (\%) & Latency & Success $\uparrow$ & TTS Mean $\downarrow$ & TTS SD \\
\midrule
C0 & 0 & 0 & 0.013 & 159.1 & 23.5 \\
C0 & 0 & 1 & 0.018 & 146.9 & 34.2 \\
C0 & 0 & 3 & 0.018 & 157.0 & 30.8 \\
C0 & 10 & 0 & 0.016 & 157.7 & 39.9 \\
C0 & 10 & 1 & 0.006 & 138.5 & 46.3 \\
C0 & 10 & 3 & 0.010 & 181.8 & 17.0 \\
C0 & 30 & 0 & 0.021 & 167.3 & 32.9 \\
C0 & 30 & 1 & 0.018 & 159.9 & 28.5 \\
C0 & 30 & 3 & 0.019 & 151.3 & 33.4 \\
\addlinespace[4pt]
C1 & 0 & 0 & 0.336 & \textbf{97.6} & 30.0 \\
C1 & 0 & 1 & 0.391 & 102.0 & 32.5 \\
C1 & 0 & 3 & 0.349 & 102.6 & 33.2 \\
C1 & 10 & 0 & 0.408 & 98.6 & 30.3 \\
C1 & 10 & 1 & 0.410 & 98.4 & 29.7 \\
C1 & 10 & 3 & 0.359 & 103.6 & 32.8 \\
C1 & 30 & 0 & 0.395 & 99.1 & 32.5 \\
C1 & 30 & 1 & 0.389 & 104.5 & 35.3 \\
C1 & 30 & 3 & 0.372 & 103.5 & 35.7 \\
\addlinespace[4pt]
C2 & 0 & 0 & 0.616 & 106.7 & 34.4 \\
C2 & 0 & 1 & 0.614 & 104.1 & 33.7 \\
C2 & 0 & 3 & 0.582 & 106.4 & 33.4 \\
C2 & 10 & 0 & 0.628 & 105.6 & 34.9 \\
C2 & 10 & 1 & 0.614 & 104.3 & 34.3 \\
C2 & 10 & 3 & 0.616 & 105.7 & 33.9 \\
C2 & 30 & 0 & 0.594 & 105.1 & 34.1 \\
C2 & 30 & 1 & 0.608 & 106.0 & 33.1 \\
C2 & 30 & 3 & 0.596 & 108.5 & 31.6 \\
\addlinespace[4pt]
C3 & 0 & 0 & 0.692 & 112.0 & 35.0 \\
C3 & 0 & 1 & 0.701 & 114.2 & 36.1 \\
C3 & 0 & 3 & 0.685 & 112.0 & 34.0 \\
C3 & 10 & 0 & 0.706 & 113.5 & 35.2 \\
C3 & 10 & 1 & \textbf{0.711} & 111.2 & 35.3 \\
C3 & 10 & 3 & 0.677 & 114.1 & 34.5 \\
C3 & 30 & 0 & 0.683 & 110.1 & 34.1 \\
C3 & 30 & 1 & 0.676 & 116.2 & 35.4 \\
C3 & 30 & 3 & 0.675 & 116.7 & 34.8 \\
\bottomrule
\end{tabular}
\end{table}

\begin{table}
\centering
\caption{Epistemic Alignment, $k=2$ agents required. JSD = mean pairwise Jensen-Shannon divergence (nats) at episode end. Alignment = mean probability mass on true target cell at episode end. Values are means across 1,000 episodes. Bold indicates best value per metric.}
\label{tab:appendix_epistemic_a}
\small
\begin{tabular}{llrrrrrrr}
\toprule
Protocol & Loss (\%) & Latency & JSD Mean & JSD SD & Align Mean $\uparrow$ & Align SD \\
\midrule
C0 & 0 & 0 & 0.4188 & 0.1013 & 0.291 & 0.188 \\
C0 & 0 & 1 & 0.4171 & 0.1031 & 0.296 & 0.192 \\
C0 & 0 & 3 & 0.4214 & 0.1007 & 0.302 & 0.190 \\
C0 & 10 & 0 & 0.4184 & 0.1017 & 0.294 & 0.188 \\
C0 & 10 & 1 & 0.4229 & 0.0989 & 0.303 & 0.189 \\
C0 & 10 & 3 & 0.4198 & 0.1006 & 0.300 & 0.190 \\
C0 & 30 & 0 & 0.4228 & 0.0989 & 0.303 & 0.189 \\
C0 & 30 & 1 & 0.4183 & 0.1020 & 0.295 & 0.190 \\
C0 & 30 & 3 & 0.4230 & 0.0993 & 0.303 & 0.187 \\
\addlinespace[4pt]
C1 & 0 & 0 & 0.0037 & 0.0085 & 0.328 & 0.436 \\
C1 & 0 & 1 & \textbf{0.0029} & 0.0112 & 0.321 & 0.428 \\
C1 & 0 & 3 & 0.0034 & 0.0113 & 0.297 & 0.414 \\
C1 & 10 & 0 & 0.0044 & 0.0221 & 0.391 & 0.452 \\
C1 & 10 & 1 & 0.0067 & 0.0232 & 0.350 & 0.439 \\
C1 & 10 & 3 & 0.0104 & 0.0343 & 0.313 & 0.420 \\
C1 & 30 & 0 & 0.0219 & 0.0565 & 0.368 & 0.451 \\
C1 & 30 & 1 & 0.0329 & 0.0726 & 0.385 & 0.447 \\
C1 & 30 & 3 & 0.0601 & 0.1009 & 0.333 & 0.421 \\
\addlinespace[4pt]
C2 & 0 & 0 & 0.0060 & 0.0090 & 0.659 & 0.470 \\
C2 & 0 & 1 & 0.0068 & 0.0171 & 0.646 & 0.469 \\
C2 & 0 & 3 & 0.0119 & 0.0453 & 0.607 & 0.471 \\
C2 & 10 & 0 & 0.0080 & 0.0150 & 0.660 & 0.468 \\
C2 & 10 & 1 & 0.0106 & 0.0242 & 0.621 & 0.472 \\
C2 & 10 & 3 & 0.0200 & 0.0606 & 0.629 & 0.456 \\
C2 & 30 & 0 & 0.0243 & 0.0501 & 0.642 & 0.460 \\
C2 & 30 & 1 & 0.0310 & 0.0582 & 0.601 & 0.465 \\
C2 & 30 & 3 & 0.0534 & 0.1028 & 0.584 & 0.450 \\
\addlinespace[4pt]
C3 & 0 & 0 & 0.0443 & 0.0697 & 0.704 & 0.430 \\
C3 & 0 & 1 & 0.0480 & 0.0689 & \textbf{0.722} & 0.427 \\
C3 & 0 & 3 & 0.0781 & 0.0587 & 0.645 & 0.415 \\
C3 & 10 & 0 & 0.0477 & 0.0694 & 0.692 & 0.429 \\
C3 & 10 & 1 & 0.0553 & 0.0664 & 0.701 & 0.429 \\
C3 & 10 & 3 & 0.0855 & 0.0669 & 0.659 & 0.394 \\
C3 & 30 & 0 & 0.0699 & 0.0858 & 0.678 & 0.409 \\
C3 & 30 & 1 & 0.0850 & 0.0915 & 0.675 & 0.410 \\
C3 & 30 & 3 & 0.1367 & 0.1109 & 0.621 & 0.368 \\
\bottomrule
\end{tabular}
\end{table}

\begin{table}
\centering
\caption{Epistemic Alignment, $k=3$ agents required. JSD = mean pairwise Jensen-Shannon divergence (nats) at episode end. Alignment = mean probability mass on true target cell at episode end. Values are means across 1,000 episodes. Bold indicates best value per metric.}
\label{tab:appendix_epistemic_b}
\small
\begin{tabular}{llrrrrrrr}
\toprule
Protocol & Loss (\%) & Latency & JSD Mean & JSD SD & Align Mean $\uparrow$ & Align SD \\
\midrule
C0 & 0 & 0 & 0.4121 & 0.1011 & 0.344 & 0.233 \\
C0 & 0 & 1 & 0.4109 & 0.1043 & 0.335 & 0.231 \\
C0 & 0 & 3 & 0.4121 & 0.1000 & 0.333 & 0.229 \\
C0 & 10 & 0 & 0.4068 & 0.1044 & 0.324 & 0.233 \\
C0 & 10 & 1 & 0.4056 & 0.1052 & 0.326 & 0.235 \\
C0 & 10 & 3 & 0.4110 & 0.1040 & 0.339 & 0.232 \\
C0 & 30 & 0 & 0.4148 & 0.1006 & 0.330 & 0.224 \\
C0 & 30 & 1 & 0.4044 & 0.1054 & 0.332 & 0.239 \\
C0 & 30 & 3 & 0.4079 & 0.1041 & 0.329 & 0.233 \\
\addlinespace[4pt]
C1 & 0 & 0 & 0.0033 & 0.0090 & 0.313 & 0.434 \\
C1 & 0 & 1 & \textbf{0.0029} & 0.0130 & 0.349 & 0.442 \\
C1 & 0 & 3 & 0.0031 & 0.0111 & 0.310 & 0.425 \\
C1 & 10 & 0 & 0.0046 & 0.0229 & 0.369 & 0.450 \\
C1 & 10 & 1 & 0.0044 & 0.0179 & 0.346 & 0.443 \\
C1 & 10 & 3 & 0.0067 & 0.0252 & 0.335 & 0.434 \\
C1 & 30 & 0 & 0.0161 & 0.0422 & 0.367 & 0.461 \\
C1 & 30 & 1 & 0.0241 & 0.0599 & 0.381 & 0.461 \\
C1 & 30 & 3 & 0.0383 & 0.0688 & 0.356 & 0.451 \\
\addlinespace[4pt]
C2 & 0 & 0 & 0.0059 & 0.0089 & 0.662 & 0.472 \\
C2 & 0 & 1 & 0.0063 & 0.0135 & 0.643 & 0.477 \\
C2 & 0 & 3 & 0.0088 & 0.0302 & 0.612 & 0.482 \\
C2 & 10 & 0 & 0.0082 & 0.0182 & 0.668 & 0.468 \\
C2 & 10 & 1 & 0.0081 & 0.0183 & 0.657 & 0.473 \\
C2 & 10 & 3 & 0.0111 & 0.0357 & 0.644 & 0.470 \\
C2 & 30 & 0 & 0.0148 & 0.0327 & 0.682 & 0.462 \\
C2 & 30 & 1 & 0.0167 & 0.0381 & 0.646 & 0.473 \\
C2 & 30 & 3 & 0.0218 & 0.0509 & 0.642 & 0.468 \\
\addlinespace[4pt]
C3 & 0 & 0 & 0.0435 & 0.0702 & 0.720 & 0.441 \\
C3 & 0 & 1 & 0.0444 & 0.0702 & \textbf{0.723} & 0.442 \\
C3 & 0 & 3 & 0.0509 & 0.0736 & 0.722 & 0.434 \\
C3 & 10 & 0 & 0.0450 & 0.0712 & 0.722 & 0.438 \\
C3 & 10 & 1 & 0.0477 & 0.0733 & 0.720 & 0.443 \\
C3 & 10 & 3 & 0.0551 & 0.0733 & 0.714 & 0.434 \\
C3 & 30 & 0 & 0.0588 & 0.0824 & 0.706 & 0.433 \\
C3 & 30 & 1 & 0.0612 & 0.0863 & 0.711 & 0.433 \\
C3 & 30 & 3 & 0.0788 & 0.0944 & 0.692 & 0.426 \\
\bottomrule
\end{tabular}
\end{table}

\begin{table}
\centering
\caption{Epistemic Alignment, $k=4$ agents required. JSD = mean pairwise Jensen-Shannon divergence (nats) at episode end. Alignment = mean probability mass on true target cell at episode end. Values are means across 1,000 episodes. Bold indicates best value per metric.}
\label{tab:appendix_epistemic_c}
\small
\begin{tabular}{llrrrrrrr}
\toprule
Protocol & Loss (\%) & Latency & JSD Mean & JSD SD & Align Mean $\uparrow$ & Align SD \\
\midrule
C0 & 0 & 0 & 0.4065 & 0.1129 & 0.337 & 0.238 \\
C0 & 0 & 1 & 0.4025 & 0.1168 & 0.335 & 0.243 \\
C0 & 0 & 3 & 0.4048 & 0.1163 & 0.331 & 0.238 \\
C0 & 10 & 0 & 0.4059 & 0.1148 & 0.345 & 0.242 \\
C0 & 10 & 1 & 0.4058 & 0.1091 & 0.322 & 0.233 \\
C0 & 10 & 3 & 0.4013 & 0.1141 & 0.326 & 0.240 \\
C0 & 30 & 0 & 0.4078 & 0.1159 & 0.342 & 0.238 \\
C0 & 30 & 1 & 0.4038 & 0.1176 & 0.337 & 0.242 \\
C0 & 30 & 3 & 0.3999 & 0.1184 & 0.332 & 0.245 \\
\addlinespace[4pt]
C1 & 0 & 0 & 0.0034 & 0.0097 & 0.313 & 0.435 \\
C1 & 0 & 1 & 0.0020 & 0.0100 & 0.363 & 0.449 \\
C1 & 0 & 3 & \textbf{0.0014} & 0.0063 & 0.316 & 0.434 \\
C1 & 10 & 0 & 0.0038 & 0.0215 & 0.378 & 0.454 \\
C1 & 10 & 1 & 0.0033 & 0.0165 & 0.381 & 0.454 \\
C1 & 10 & 3 & 0.0059 & 0.0263 & 0.323 & 0.436 \\
C1 & 30 & 0 & 0.0162 & 0.0443 & 0.382 & 0.467 \\
C1 & 30 & 1 & 0.0172 & 0.0452 & 0.373 & 0.466 \\
C1 & 30 & 3 & 0.0290 & 0.0585 & 0.363 & 0.461 \\
\addlinespace[4pt]
C2 & 0 & 0 & 0.0060 & 0.0090 & 0.654 & 0.475 \\
C2 & 0 & 1 & 0.0072 & 0.0236 & 0.648 & 0.476 \\
C2 & 0 & 3 & 0.0081 & 0.0281 & 0.614 & 0.484 \\
C2 & 10 & 0 & 0.0073 & 0.0109 & 0.661 & 0.473 \\
C2 & 10 & 1 & 0.0080 & 0.0170 & 0.648 & 0.476 \\
C2 & 10 & 3 & 0.0089 & 0.0265 & 0.655 & 0.472 \\
C2 & 30 & 0 & 0.0152 & 0.0304 & 0.640 & 0.478 \\
C2 & 30 & 1 & 0.0136 & 0.0265 & 0.654 & 0.474 \\
C2 & 30 & 3 & 0.0169 & 0.0410 & 0.635 & 0.477 \\
\addlinespace[4pt]
C3 & 0 & 0 & 0.0426 & 0.0693 & 0.727 & 0.442 \\
C3 & 0 & 1 & 0.0426 & 0.0714 & 0.733 & 0.439 \\
C3 & 0 & 3 & 0.0453 & 0.0741 & 0.725 & 0.443 \\
C3 & 10 & 0 & 0.0408 & 0.0694 & \textbf{0.747} & 0.432 \\
C3 & 10 & 1 & 0.0415 & 0.0709 & 0.746 & 0.432 \\
C3 & 10 & 3 & 0.0484 & 0.0773 & 0.718 & 0.444 \\
C3 & 30 & 0 & 0.0527 & 0.0889 & 0.730 & 0.436 \\
C3 & 30 & 1 & 0.0538 & 0.0897 & 0.727 & 0.435 \\
C3 & 30 & 3 & 0.0542 & 0.0930 & 0.739 & 0.427 \\
\bottomrule
\end{tabular}
\end{table}

\begin{table}
\centering
\caption{MEH and Communication Cost, $k=2$ agents required. MEH = malignant epistemic herding rate among failed episodes (JSD $< 0.1$ nats). Msgs = mean messages sent per episode. APB = alignment per byte ($\times 10^{-6}$). C\textsubscript{0} excluded from communication metrics (no transmission). Values are means across 1,000 episodes. Bold indicates best value per metric.}
\label{tab:appendix_cost_a}
\small
\begin{tabular}{llrrrrrr}
\toprule
Protocol & Loss (\%) & Latency & MEH Rate $\downarrow$ & Msgs $\downarrow$ & Bytes $\downarrow$ & APB $\uparrow$ \\
\midrule
C0 & 0 & 0 & \textbf{0.000} & --- & --- & --- \\
C0 & 0 & 1 & 0.000 & --- & --- & --- \\
C0 & 0 & 3 & 0.000 & --- & --- & --- \\
C0 & 10 & 0 & 0.000 & --- & --- & --- \\
C0 & 10 & 1 & 0.000 & --- & --- & --- \\
C0 & 10 & 3 & 0.000 & --- & --- & --- \\
C0 & 30 & 0 & 0.000 & --- & --- & --- \\
C0 & 30 & 1 & 0.000 & --- & --- & --- \\
C0 & 30 & 3 & 0.000 & --- & --- & --- \\
\addlinespace[4pt]
C1 & 0 & 0 & 0.639 & 1894.9 & 13264.3 & 24.7 \\
C1 & 0 & 1 & 0.635 & 1880.1 & 13161.0 & 24.4 \\
C1 & 0 & 3 & 0.649 & 1897.6 & 13283.3 & 22.4 \\
C1 & 10 & 0 & 0.565 & 1788.5 & 12519.4 & 31.3 \\
C1 & 10 & 1 & 0.597 & 1834.8 & 12843.6 & 27.3 \\
C1 & 10 & 3 & 0.614 & 1873.5 & 13114.4 & 23.9 \\
C1 & 30 & 0 & 0.576 & 1845.7 & 12920.2 & 28.5 \\
C1 & 30 & 1 & 0.529 & 1821.6 & 12750.9 & 30.2 \\
C1 & 30 & 3 & 0.516 & 1865.2 & 13056.1 & 25.5 \\
\addlinespace[4pt]
C2 & 0 & 0 & 0.358 & 1509.5 & 52833.1 & 12.5 \\
C2 & 0 & 1 & 0.360 & 1529.0 & 53516.4 & 12.1 \\
C2 & 0 & 3 & 0.384 & 1587.0 & 55544.6 & 10.9 \\
C2 & 10 & 0 & 0.350 & 1498.7 & 52455.5 & 12.6 \\
C2 & 10 & 1 & 0.380 & 1534.5 & 53707.5 & 11.6 \\
C2 & 10 & 3 & 0.362 & 1569.6 & 54936.8 & 11.4 \\
C2 & 30 & 0 & 0.348 & 1532.3 & 53629.0 & 12.0 \\
C2 & 30 & 1 & 0.384 & 1572.6 & 55042.7 & 10.9 \\
C2 & 30 & 3 & 0.369 & 1569.4 & 54930.5 & 10.6 \\
\addlinespace[4pt]
C3 & 0 & 0 & 0.016 & 20.4 & 714.7 & 985.0 \\
C3 & 0 & 1 & 0.018 & 17.2 & 602.5 & \textbf{1199.1} \\
C3 & 0 & 3 & 0.014 & \textbf{15.9} & \textbf{554.8} & 1163.3 \\
C3 & 10 & 0 & 0.019 & 21.6 & 756.1 & 915.2 \\
C3 & 10 & 1 & 0.016 & 18.8 & 656.5 & 1067.5 \\
C3 & 10 & 3 & 0.016 & 17.6 & 615.6 & 1069.9 \\
C3 & 30 & 0 & 0.019 & 23.5 & 821.1 & 826.3 \\
C3 & 30 & 1 & 0.011 & 20.9 & 731.4 & 922.8 \\
C3 & 30 & 3 & 0.010 & 19.3 & 675.1 & 920.0 \\
\bottomrule
\end{tabular}
\end{table}

\begin{table}
\centering
\caption{MEH and Communication Cost, $k=3$ agents required. MEH = malignant epistemic herding rate among failed episodes (JSD $< 0.1$ nats). Msgs = mean messages sent per episode. APB = alignment per byte ($\times 10^{-6}$). C\textsubscript{0} excluded from communication metrics (no transmission). Values are means across 1,000 episodes. Bold indicates best value per metric.}
\label{tab:appendix_cost_b}
\small
\begin{tabular}{llrrrrrr}
\toprule
Protocol & Loss (\%) & Latency & MEH Rate $\downarrow$ & Msgs $\downarrow$ & Bytes $\downarrow$ & APB $\uparrow$ \\
\midrule
C0 & 0 & 0 & \textbf{0.000} & --- & --- & --- \\
C0 & 0 & 1 & 0.000 & --- & --- & --- \\
C0 & 0 & 3 & 0.000 & --- & --- & --- \\
C0 & 10 & 0 & 0.000 & --- & --- & --- \\
C0 & 10 & 1 & 0.000 & --- & --- & --- \\
C0 & 10 & 3 & 0.000 & --- & --- & --- \\
C0 & 30 & 0 & 0.000 & --- & --- & --- \\
C0 & 30 & 1 & 0.000 & --- & --- & --- \\
C0 & 30 & 3 & 0.000 & --- & --- & --- \\
\addlinespace[4pt]
C1 & 0 & 0 & 0.661 & 1943.2 & 13602.7 & 23.0 \\
C1 & 0 & 1 & 0.612 & 1897.7 & 13283.8 & 26.3 \\
C1 & 0 & 3 & 0.647 & 1920.3 & 13442.0 & 23.1 \\
C1 & 10 & 0 & 0.592 & 1855.6 & 12989.2 & 28.4 \\
C1 & 10 & 1 & 0.617 & 1900.6 & 13304.4 & 26.0 \\
C1 & 10 & 3 & 0.612 & 1909.0 & 13363.1 & 25.1 \\
C1 & 30 & 0 & 0.590 & 1887.3 & 13211.4 & 27.8 \\
C1 & 30 & 1 & 0.552 & 1879.9 & 13159.2 & 29.0 \\
C1 & 30 & 3 & 0.519 & 1912.7 & 13388.9 & 26.6 \\
\addlinespace[4pt]
C2 & 0 & 0 & 0.371 & 1607.5 & 56264.0 & 11.8 \\
C2 & 0 & 1 & 0.378 & 1616.5 & 56576.9 & 11.4 \\
C2 & 0 & 3 & 0.402 & 1664.5 & 58259.0 & 10.5 \\
C2 & 10 & 0 & 0.363 & 1585.7 & 55500.9 & 12.0 \\
C2 & 10 & 1 & 0.362 & 1596.2 & 55868.0 & 11.8 \\
C2 & 10 & 3 & 0.367 & 1628.1 & 56984.8 & 11.3 \\
C2 & 30 & 0 & 0.351 & 1614.8 & 56518.1 & 12.1 \\
C2 & 30 & 1 & 0.376 & 1666.1 & 58312.0 & 11.1 \\
C2 & 30 & 3 & 0.368 & 1654.7 & 57915.9 & 11.1 \\
\addlinespace[4pt]
C3 & 0 & 0 & 0.031 & 24.0 & 841.7 & 855.9 \\
C3 & 0 & 1 & 0.023 & \textbf{22.4} & \textbf{785.0} & \textbf{920.9} \\
C3 & 0 & 3 & 0.031 & 22.4 & 785.0 & 920.0 \\
C3 & 10 & 0 & 0.029 & 26.4 & 924.9 & 780.3 \\
C3 & 10 & 1 & 0.022 & 24.5 & 857.7 & 839.6 \\
C3 & 10 & 3 & 0.019 & 23.8 & 833.4 & 856.4 \\
C3 & 30 & 0 & 0.028 & 27.8 & 972.4 & 726.4 \\
C3 & 30 & 1 & 0.031 & 26.5 & 926.8 & 766.8 \\
C3 & 30 & 3 & 0.018 & 25.1 & 878.6 & 787.5 \\
\bottomrule
\end{tabular}
\end{table}

\begin{table}
\centering
\caption{MEH and Communication Cost, $k=4$ agents required. MEH = malignant epistemic herding rate among failed episodes (JSD $< 0.1$ nats). Msgs = mean messages sent per episode. APB = alignment per byte ($\times 10^{-6}$). C\textsubscript{0} excluded from communication metrics (no transmission). Values are means across 1,000 episodes. Bold indicates best value per metric.}
\label{tab:appendix_cost_c}
\small
\begin{tabular}{llrrrrrr}
\toprule
Protocol & Loss (\%) & Latency & MEH Rate $\downarrow$ & Msgs $\downarrow$ & Bytes $\downarrow$ & APB $\uparrow$ \\
\midrule
C0 & 0 & 0 & \textbf{0.000} & --- & --- & --- \\
C0 & 0 & 1 & 0.000 & --- & --- & --- \\
C0 & 0 & 3 & 0.000 & --- & --- & --- \\
C0 & 10 & 0 & 0.000 & --- & --- & --- \\
C0 & 10 & 1 & 0.000 & --- & --- & --- \\
C0 & 10 & 3 & 0.000 & --- & --- & --- \\
C0 & 30 & 0 & 0.000 & --- & --- & --- \\
C0 & 30 & 1 & 0.000 & --- & --- & --- \\
C0 & 30 & 3 & 0.000 & --- & --- & --- \\
\addlinespace[4pt]
C1 & 0 & 0 & 0.664 & 1991.3 & 13939.1 & 22.4 \\
C1 & 0 & 1 & 0.606 & 1944.8 & 13613.7 & 26.7 \\
C1 & 0 & 3 & 0.651 & 1996.3 & 13974.0 & 22.6 \\
C1 & 10 & 0 & 0.584 & 1908.4 & 13358.6 & 28.3 \\
C1 & 10 & 1 & 0.581 & 1905.0 & 13335.2 & 28.5 \\
C1 & 10 & 3 & 0.630 & 1988.8 & 13921.7 & 23.2 \\
C1 & 30 & 0 & 0.575 & 1926.4 & 13485.0 & 28.3 \\
C1 & 30 & 1 & 0.576 & 1958.9 & 13712.1 & 27.2 \\
C1 & 30 & 3 & 0.549 & 1973.5 & 13814.3 & 26.3 \\
\addlinespace[4pt]
C2 & 0 & 0 & 0.384 & 1717.7 & 60118.8 & 10.9 \\
C2 & 0 & 1 & 0.383 & 1701.0 & 59534.6 & 10.9 \\
C2 & 0 & 3 & 0.412 & 1753.6 & 61376.3 & 10.0 \\
C2 & 10 & 0 & 0.372 & 1696.4 & 59374.6 & 11.1 \\
C2 & 10 & 1 & 0.384 & 1702.5 & 59586.2 & 10.9 \\
C2 & 10 & 3 & 0.378 & 1710.6 & 59871.8 & 10.9 \\
C2 & 30 & 0 & 0.401 & 1730.9 & 60580.4 & 10.6 \\
C2 & 30 & 1 & 0.389 & 1721.4 & 60250.7 & 10.9 \\
C2 & 30 & 3 & 0.392 & 1752.4 & 61335.1 & 10.3 \\
\addlinespace[4pt]
C3 & 0 & 0 & 0.038 & \textbf{24.3} & \textbf{848.8} & \textbf{856.5} \\
C3 & 0 & 1 & 0.034 & 24.7 & 864.1 & 848.4 \\
C3 & 0 & 3 & 0.043 & 24.3 & 850.7 & 852.8 \\
C3 & 10 & 0 & 0.042 & 27.3 & 955.6 & 781.3 \\
C3 & 10 & 1 & 0.037 & 27.3 & 956.0 & 780.7 \\
C3 & 10 & 3 & 0.044 & 26.2 & 918.6 & 782.1 \\
C3 & 30 & 0 & 0.039 & 28.8 & 1008.1 & 723.9 \\
C3 & 30 & 1 & 0.041 & 28.6 & 1002.4 & 725.6 \\
C3 & 30 & 3 & 0.055 & 28.2 & 986.6 & 749.2 \\
\bottomrule
\end{tabular}
\end{table}

\section{Scaling Analysis}
Table~\ref{tab:appendix_scaling} reports results across grid sizes $\{25\times25, 50\times50, 75\times75, 100\times100\}$ with episode budgets scaled proportionally to grid side length. All conditions use $k=3$, no packet loss, and no latency.

\begin{table}[htbp]
\centering
\caption{Scaling Analysis Across Grid Sizes ($k=3$, no packet loss, no latency). Each condition uses 1{,}000 episodes (500 for $100 \times 100$). Episode budgets scale proportionally with grid side length. Bold indicates best value per metric within each grid size block.}
\label{tab:appendix_scaling}
\small
\begin{tabular}{llrrrrrr}
\toprule
Protocol & Grid & $T$ & Success $\uparrow$ & MEH Rate $\downarrow$ & Align $\uparrow$ & JSD\\
\midrule
\multicolumn{7}{l}{\textit{Grid: $25\times 25$, $T=50$}} \\
\addlinespace[2pt]
C0 & $25\times25$ & 50 & 0.086 & \textbf{0.000} & 0.326 & 0.4122 \\
C1 & $25\times25$ & 50 & 0.472 & 0.528 & 0.473 & \textbf{0.0038} \\
C2 & $25\times25$ & 50 & 0.672 & 0.328 & \textbf{0.750} & 0.0108 \\
C3 & $25\times25$ & 50 & \textbf{0.677} & 0.079 & 0.745 & 0.0423 \\
\midrule
\multicolumn{7}{l}{\textit{Grid: $50\times 50$, $T=200$}} \\
\addlinespace[2pt]
C0 & $50\times50$ & 200 & 0.134 & \textbf{0.000} & 0.341 & 0.4088 \\
C1 & $50\times50$ & 200 & 0.347 & 0.653 & 0.319 & \textbf{0.0030} \\
C2 & $50\times50$ & 200 & 0.660 & 0.340 & 0.691 & 0.0052 \\
C3 & $50\times50$ & 200 & \textbf{0.705} & 0.024 & \textbf{0.723} & 0.0431 \\
\midrule
\multicolumn{7}{l}{\textit{Grid: $75\times 75$, $T=350$}} \\
\addlinespace[2pt]
C0 & $75\times75$ & 350 & 0.075 & \textbf{0.000} & 0.273 & 0.3723 \\
C1 & $75\times75$ & 350 & 0.230 & 0.770 & 0.211 & \textbf{0.0036} \\
C2 & $75\times75$ & 350 & 0.527 & 0.473 & 0.541 & 0.0054 \\
C3 & $75\times75$ & 350 & \textbf{0.615} & 0.026 & \textbf{0.634} & 0.0491 \\
\midrule
\multicolumn{7}{l}{\textit{Grid: $100\times 100$, $T=500$}} \\
\addlinespace[2pt]
C0 & $100\times100$ & 500 & 0.026 & \textbf{0.000} & 0.210 & 0.3269 \\
C1 & $100\times100$ & 500 & 0.158 & 0.842 & 0.145 & \textbf{0.0033} \\
C2 & $100\times100$ & 500 & 0.451 & 0.549 & 0.472 & 0.0049 \\
C3 & $100\times100$ & 500 & \textbf{0.582} & 0.035 & \textbf{0.608} & 0.0478 \\
\bottomrule
\end{tabular}
\end{table}

\section{Entropy-Delta Sensitivity Across Coordination Thresholds}
Table~\ref{tab:appendix_delta} extends Table~5 from the main paper to include sensitivity results for $k \in \{2, 3, 4\}$. The selected operating point $\theta=0.20$ minimizes MEH within the low-bandwidth plateau across all coordination thresholds.

\begin{table}[htbp]
\centering
\caption{Sensitivity of Entropy-Delta Gating (C\textsubscript{3}) to Threshold $\theta$ Across All Coordination Requirements ($50\times50$ grid, no packet loss, no latency, 1{,}000 episodes). $\theta=0$ corresponds to ungated epistemic communication (equivalent to C\textsubscript{2}). Bold indicates best value per column; underline marks the selected operating point ($\theta=0.20$).}
\label{tab:appendix_delta}
\small
\begin{tabular}{lrrrrrrr}
\toprule
$\theta$ & Success $\uparrow$ & Time $\downarrow$ & MEH Rate $\downarrow$ & JSD & Align $\uparrow$ & Msgs $\downarrow$ & APB $\uparrow$ \\
\midrule
\multicolumn{8}{l}{\textit{Coordination threshold: $k=2$ agents}} \\
\addlinespace[2pt]
0.00 & 0.617 & \textbf{84.6} & 0.383 & 0.0061 & 0.631 & 1552.7 & 11.6 \\
0.05 & 0.641 & 86.0 & 0.234 & 0.0338 & 0.648 & 47.9 & 386.7 \\
0.10 & 0.675 & 85.5 & 0.085 & 0.0375 & 0.670 & 29.4 & 650.8 \\
0.15 & 0.650 & 88.9 & 0.039 & 0.0456 & 0.649 & 23.4 & 794.3 \\
\underline{0.20} & \underline{0.705} & \underline{89.9} & \underline{0.022} & \underline{0.0449} & \underline{0.702} & \underline{20.3} & \underline{988.3} \\
0.25 & 0.730 & 91.0 & \textbf{0.010} & 0.0437 & 0.716 & 19.5 & 1050.5 \\
0.30 & 0.716 & 93.7 & 0.022 & \textbf{0.0469} & 0.713 & 18.2 & 1120.1 \\
0.35 & \textbf{0.781} & 94.0 & 0.023 & 0.0387 & \textbf{0.777} & 18.0 & 1229.6 \\
0.40 & 0.756 & 93.8 & 0.023 & 0.0464 & 0.753 & \textbf{16.9} & \textbf{1274.6} \\
\midrule
\multicolumn{8}{l}{\textit{Coordination threshold: $k=3$ agents}} \\
\addlinespace[2pt]
0.00 & 0.646 & \textbf{92.2} & 0.354 & 0.0057 & 0.672 & 1572.0 & 12.2 \\
0.05 & 0.606 & 95.8 & 0.225 & 0.0366 & 0.627 & 51.8 & 345.5 \\
0.10 & 0.659 & 95.5 & 0.083 & 0.0381 & 0.675 & 33.1 & 582.0 \\
0.15 & 0.674 & 97.3 & 0.047 & 0.0395 & 0.701 & 27.1 & 739.7 \\
\underline{0.20} & \underline{0.687} & \underline{100.1} & \underline{\textbf{0.025}} & \underline{\textbf{0.0452}} & \underline{0.704} & \underline{23.9} & \underline{843.7} \\
0.25 & 0.709 & 101.6 & 0.032 & 0.0425 & 0.734 & 23.5 & 892.4 \\
0.30 & 0.738 & 101.0 & 0.039 & 0.0389 & 0.768 & 22.7 & 965.7 \\
0.35 & \textbf{0.758} & 104.5 & 0.049 & 0.0371 & \textbf{0.798} & 22.2 & 1029.6 \\
0.40 & 0.737 & 103.8 & 0.046 & 0.0449 & 0.777 & \textbf{19.2} & \textbf{1158.3} \\
\midrule
\multicolumn{8}{l}{\textit{Coordination threshold: $k=4$ agents}} \\
\addlinespace[2pt]
0.00 & 0.625 & \textbf{104.2} & 0.375 & 0.0058 & 0.669 & 1689.1 & 11.3 \\
0.05 & 0.599 & 108.9 & 0.269 & 0.0350 & 0.631 & 54.1 & 333.4 \\
0.10 & 0.626 & 107.8 & 0.108 & 0.0399 & 0.663 & 34.7 & 545.5 \\
0.15 & 0.650 & 110.4 & 0.049 & 0.0412 & 0.688 & 27.0 & 727.9 \\
\underline{0.20} & \underline{0.694} & \underline{114.4} & \underline{\textbf{0.034}} & \underline{\textbf{0.0437}} & \underline{0.726} & \underline{24.4} & \underline{848.4} \\
0.25 & 0.721 & 115.9 & 0.051 & 0.0373 & 0.770 & 24.1 & 911.1 \\
0.30 & 0.724 & 111.8 & 0.059 & 0.0383 & 0.779 & 22.9 & 970.8 \\
0.35 & \textbf{0.727} & 114.6 & 0.072 & 0.0386 & 0.794 & 22.1 & 1028.4 \\
0.40 & 0.721 & 114.7 & 0.080 & 0.0409 & \textbf{0.797} & \textbf{21.5} & \textbf{1061.1} \\
\bottomrule
\end{tabular}
\end{table}

\end{appendices}

\end{document}